\documentclass[apsprl,twocolumn,longbibliography]{revtex4-2}

\usepackage{physics}
\usepackage{graphicx}
\usepackage{amsmath}
\usepackage[usenamesdvipsnames]{xcolor}
\usepackage[colorlinks=truelinkcolor=blue,urlcolor=blue,citecolor=blue]{hyperref}
\usepackage{braket}
\usepackage{amssymb}
\usepackage{commath}
\usepackage{csquotes}
\usepackage{dcolumn}
\usepackage{bm}
\usepackage{epstopdf}
\usepackage{lipsum}
\usepackage{subfigure}
\usepackage{siunitx}
\usepackage{pifont}

\begin{document}

\title{Observing non-ergodicity due to kinetic constraints in tilted Fermi-Hubbard chains}

\author{Sebastian~Scherg$^{1,2,3}$, Thomas~Kohlert$^{1,2,3}$, Pablo~Sala$^{3,4}$, Frank~Pollmann$^{3,4}$, Bharath~H.~M.$^{1,2,3}$, Immanuel~Bloch$^{1,2,3}$, Monika~Aidelsburger$^{1,3}$}

\affiliation{$^{1}$\,Fakult\"at f\"ur Physik, Ludwig-Maximilians-Universit\"at M\"unchen, 
80799 Munich, Germany}
\affiliation{$^{2}$\,Max-Planck-Institut f\"ur Quantenoptik, 
85748 Garching, Germany}
\affiliation{$^{3}$\,Munich Center for Quantum Science and Technology (MCQST), 
80799 M\"unchen, Germany}
\affiliation{$^{4}$\,Department of Physics, Technical University of Munich, 
85748 Garching, Germany}

\maketitle



\textbf{
	The thermalization of isolated quantum many-body systems is deeply related to fundamental questions of quantum information theory. While integrable or many-body localized systems display non-ergodic behavior due to extensively many conserved quantities, recent theoretical studies have identified a rich variety of more exotic phenomena in between these two extreme limits. The tilted one-dimensional Fermi-Hubbard model, which is readily accessible in experiments with ultracold atoms, emerged as an intriguing playground to study non-ergodic behavior in a clean disorder-free system. While non-ergodic behavior was established theoretically in certain limiting cases, there is no complete understanding of the complex thermalization properties of this model. In this work, we experimentally study the relaxation of an initial charge-density wave and find a remarkably long-lived initial-state memory over a wide range of parameters. Our observations are well reproduced by numerical simulations of a clean system. Using analytical calculations we further provide a detailed microscopic understanding of this behavior, which can be attributed to emergent kinetic constraints.}


Understanding the complex out-of-equilibrium dynamics of quantum many-body systems is central to a number of research areas ranging from statistical physics to quantum information  theory~\cite{gogolin_equilibration_2016, alessio_from_2016,mori_thermalization_2018}. State-of-the-art experimental platforms are now able to test novel theoretical concepts and approximate descriptions based on experimental observations. 
Important experimental results were obtained in particular with integrable~\cite{calabrese_quantum_2011} or many-body localized (MBL)~\cite{nandkishore_many-body_2015,altman_universal_2015,abanin_colloquium_2019} systems. Both phenomena emerge due to the existence of extensively many conserved quantities and have been of considerable interest, because they break the eigenstate thermalization hypothesis, which assumes that each individual eigenstate behaves locally like a thermal ensemble and is believed to hold for generic ergodic systems~\cite{deutsch_quantum_1991,srednicki_chaos_1994,rigol_thermalization_2008}.

\begin{figure}[t]
	\includegraphics[width=3.3in]{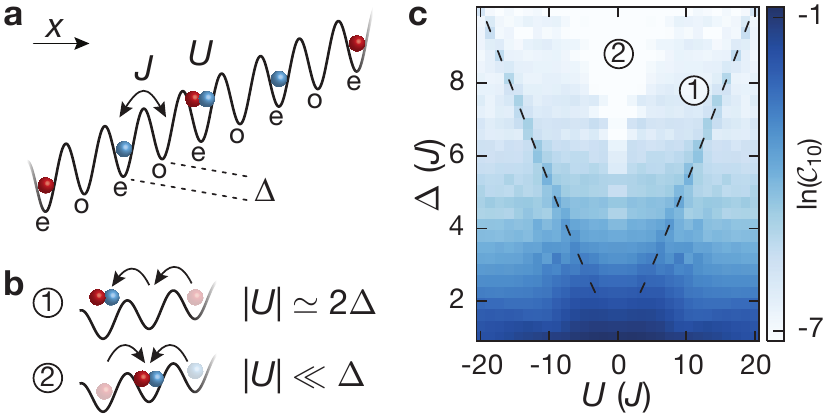}	  
	\caption{\textbf{Illustration of the experimental setup and the structure of the Hilbert space.} 
		\textbf{a} Schematic of the tilted 1D Fermi-Hubbard model (with odd \textit{o} and even \textit{e} sites) with tunneling $J$, on-site interaction $U$ and spin-dependent tilt $\Delta_\uparrow,\Delta_\downarrow$ (spin-up red, spin-down blue). 	
		\textbf{b} Dominant resonant tunneling processes for different regimes.
		\textbf{c} Finite-time connectivity $\mathcal{C}_{\epsilon}$ (for a cut-off $\epsilon=10 \%$) defined as the fraction of states that participate in the dynamics up to an evolution time $T_{\mathcal{N}}=1000\tau$ (main text, Methods). The calculation was performed for a N\'eel-ordered singlon CDW initial state, using exact diagonalization (ED) with system size $L=13$ and $\Delta_\uparrow=\Delta_\downarrow\equiv\Delta$.  
		In the large-tilt limit, $\Delta/J\rightarrow \infty$, we find emergent strongly-fragmented effective Hamiltonians for regime \ding{172} and \ding{173}~\cite{supplements}.}
	\label{fig1}
\end{figure}

In between the two extreme limits of ergodic and localizing dynamics there exists a rich variety of more complex thermalizing behavior. Models with many-body scar states, e.g., host a vanishing fraction of non-thermal eigenstates embedded within an otherwise thermal spectrum~\cite{shiraishi_systematic_2017,mondaini_comment_2018,moudgalya_exact_2018,turner_weak_2018,iadecola_quantum_2020,chattopadhyay_quantum_2020}. They exhibit a weak form of ergodicity-breaking, that strongly depends on the initial state, as has been observed with Rydberg atoms~\cite{bernien_probing_2017,turner_weak_2018,turner_quantum_2018}. 
More recently, a whole new class of models has been suggested, where the presence of only few conserved quantities, in particular dipole conservation, results in non-ergodic dynamics due to an emergent fragmentation of the Hilbert space into exponentially many disconnected subspaces~\cite{sala_ergodicity_2020,khemani_localization_2020,rakovszky_statistical_2020, moudgalya_thermalization_2019}. Fragmented models offer an alternative view on a central open question, namely if many-body localization can occur in translationally-invariant models without disorder~\cite{van_horssen_dynamics_2015,schiulaz_dynamics_2015,yao_quasi-many-body_2016,papic_many-body_2015,smith_disorder-free_2017,smith_absence_2017,brenes_many-body_2018}.

In this work we study non-ergodic behavior in the disorder-free tilted one-dimensional (1D) Fermi-Hubbard model (Fig.~\ref{fig1}a), which lies at the interface of MBL and Hilbert-space fragmentation. In the presence of additional weak disorder or harmonic confinement, theoretical studies have found characteristic MBL phenomenology, known as Stark MBL~\cite{schulz_stark_2019,nieuwenburg_bloch_2019,wu_bath-induced_2019,taylor_experimental_2020,yao_many-body_2020}.
This, however, does not hold for a clean system with pure linear potential~\cite{schulz_stark_2019,taylor_experimental_2020}. While conventional MBL predicts localization for any typical initial state, we do not expect this to hold for our system, where resonances can occur between interaction and tilt energies (regime \ding{172} in Fig.~\ref{fig1}b). 
Intriguingly, it has been predicted, that in the limit of large tilts, $\Delta\gg J,|U|$, non-ergodicity may still occur despite the absence of disorder. In this regime, the large tilt energy imposes kinetic constraints, which result in an emergent dipole conservation~\cite{nieuwenburg_bloch_2019,sala_ergodicity_2020,khemani_localization_2020, taylor_experimental_2020, moudgalya_thermalization_2019}. This emergent behavior is in fact governed by a fragmented Hamiltonian resulting in non-ergodic dynamics. 

Starting from an initial charge-density wave (CDW) of singlons (singly-occupied site), we study relaxation dynamics in the tilted 1D Fermi-Hubbard model for a large range of interaction strengths and moderate values of the tilt ($\Delta<4J$), where none of the two mechanisms described above should apply and where naively one may expect the system to thermalize~\cite{ott_collisionally_2004,strohmaier_interaction-controlled_2007}.
At short times we observe coherent dynamics due to Bloch oscillations, whose amplitude strongly depends on the Hubbard interactions. Surprisingly we find that after intermediate times and even close to resonance (regime \ding{172}), the evolution converges to a steady-state, that persists for long evolution times up to $700$ tunneling times, signaling a robust memory of the initial CDW throughout. 

Using numerical calculations we show that the observed non-ergodicity cannot be explained by the phenomenon of Stark-MBL, i.e., the robust memory is not due to experimental imperfections, such as residual harmonic confinement or disorder, and the bipartite entanglement entropy does not exhibit the characteristic behavior of MBL systems~\cite{bardarson_unbounded_2012,schulz_stark_2019} (Fig.~\ref{fig:scaling}). Hence, non-ergodicity appears to have a different origin, despite similar experimental signatures. This raises the question about the origin of the observed non-ergodicity. We construct effective Hamiltonians in two distinct regimes (\ding{172} and \ding{173}, Fig.~\ref{fig1}b) by taking the large tilt limit and find strongly-fragmented Hamiltonians in both cases (Sect.~\ref{sect:effHam}). While these models are only expected to describe the dynamics at large tilt values and for intermediate times (on the order of a few tens of tunneling times), they allow us to identify the microscopic processes that initiate dynamics at short times (Fig.~\ref{fig1}b). In both regimes these are correlated tunneling processes, which result in the formation of doublons (doubly-occupied sites),  either resonantly (regime \ding{172}) or detuned by the Hubbard interaction energy $U$ (regime \ding{173}). Higher-order terms are expected to eventually drive the system towards thermalization~\cite{sala_ergodicity_2020}. However, we are able to show that energy penalties for the second- or higher-order tunneling processes, which occur naturally in the model, render these dynamics inefficient. This results in extremely slow relaxation (Sect.~\ref{sect:effHam}), which appears stable for $>10^4$ tunneling times in our exact diagonalization studies of small systems, in agreement with our experimental observations (Fig.~\ref{sec:penalty}).

In order to characterize the dynamics across the whole parameter regime studied experimentally, we compute the finite-time connectivity of our initial CDW state $\mathcal{C}_{\epsilon}=\mathrm{dim}(\mathcal{N}_{\epsilon})/ \mathrm{dim}(\mathcal{H})$, which is defined by the fraction of states that participate in the time evolution up to a finite time $T_\mathcal{N}$; here $\mathcal{N}_\epsilon$ denotes the subspace in the complete Hilbert-space $\mathcal{H}$, which is defined, such that the residual overlap of the time-evolved state $|\psi(t)\rangle$ outside of $\mathcal{N}_\epsilon$ is at most $\epsilon$ at any time $t\leq T_{\mathcal{N}}$ (Methods). The value of $\epsilon$ is typically chosen between $1\%$-$10\%$. The finite-time connectivity can be understood as a measure of non-ergodicity, similar to the more conventional return probability or other multifractality measures~\cite{luca_ergodicity_2013}. While effective Hamiltonians can only be derived explicitly in certain limits, the numerical construction is applicable in the whole parameter regime probed in this work (Fig.~\ref{fig1}c). 
We find that the finite-time connectivity vanishes in the thermodynamic limit for all parameters, suggesting that only a small fraction of the states participates in the dynamics, signaling non-ergodic behavior. Our results suggest that the emergent kinetic constraints result in transient non-ergodic behavior across the whole parameter range studied in this work. We further show analytically that the relevant microscopic constraints in the resonant \ding{172} regime give rise to Hilbert-space fragmentation in the large tilt limit (Sect.\ref{sect:resfrag} in~\cite{supplements}).


The experimental setup consists of a degenerate Fermi gas of $50(5) \times 10^3$ $^{40}\mathrm{K}$ atoms that is prepared in an equal mixture of two spin components $\ket {\uparrow} = \ket{m_F = -7/2}$ and $\ket{\downarrow} = \ket{m_F = -9/2}$ in the $F=9/2$ ground-state hyperfine manifold. The atoms are loaded into a 3D optical lattice with lattice constant $d_s = \SI{266}{nm}$ along the $x$ direction and deep transverse lattices, with constant $d_\perp =  \SI{369}{nm}$, to isolate the 1D chains along $x$ (Methods). The central 1D chains have a length of about 290 lattice sites. The residual coupling along the transverse directions is less than $3\times10^{-4}J$. The dynamics along $x$ is described by the tilted 1D Fermi-Hubbard model

\begin{equation}
\begin{split}
\hat{H}=\sum_{i,\sigma=\uparrow,\downarrow} &\left(-J\hat{c}_{i,\sigma}^{\dagger}\hat{c}_{i+1,\sigma}+ \textrm{h.c.}  + \Delta_{\sigma} i\hat{n}_{i,\sigma} \right)\\
&+U\sum_{i}\hat{n}_{i,\uparrow}\hat{n}_{i,\downarrow} \,,
\end{split}
\label{eq:Hamiltonian}
\end{equation}

\noindent where $\hat{c}_{i \sigma}^{\dagger}$ ($\hat{c}_{i \sigma}$) is the fermionic creation (annihilation) operator and $\hat{n}_{i,\sigma}=\hat{c}_{i \sigma}^{\dagger}\hat{c}_{i \sigma}$. The on-site interaction strength $U$ is controlled by a Feshbach resonance centered at $202.1\,$G and a magnetic field gradient is used to create the tilt $\Delta_{\sigma}$, with $\Delta_\uparrow \simeq 0.9 \Delta_\downarrow$. The weak spin-dependence arises due to the different $m_{F}$ quantum numbers~\cite{supplements}. The initial state for all subsequent measurements is a CDW of singlons on even sites, which is prepared using a bichromatic optical superlattice (Sect.~\ref{sec:sequence} in~\cite{supplements}). The initial state can be described as an incoherent mixture of site-localized particles with random spin configuration (Methods). The subsequent evolution is monitored by extracting the spin-resolved imbalance $\mathcal{I}^{\sigma}=(N_{e}^{\sigma}-N_{o}^{\sigma})/N^\sigma$~\cite{imbalance}; here $N_{e(o)}^{\sigma}$ denotes the total number of spin-$\sigma$ atoms on even (odd)  sites and $N^\sigma=N_{e}^{\sigma}+N_{o}^{\sigma}$. 
A non-zero steady-state imbalance signals a memory of the initial state~\cite{temperature}, where $\mathcal{I}^\sigma(t=0)=1$.


\begin{figure*}[ht!]
	\includegraphics[width=\textwidth]{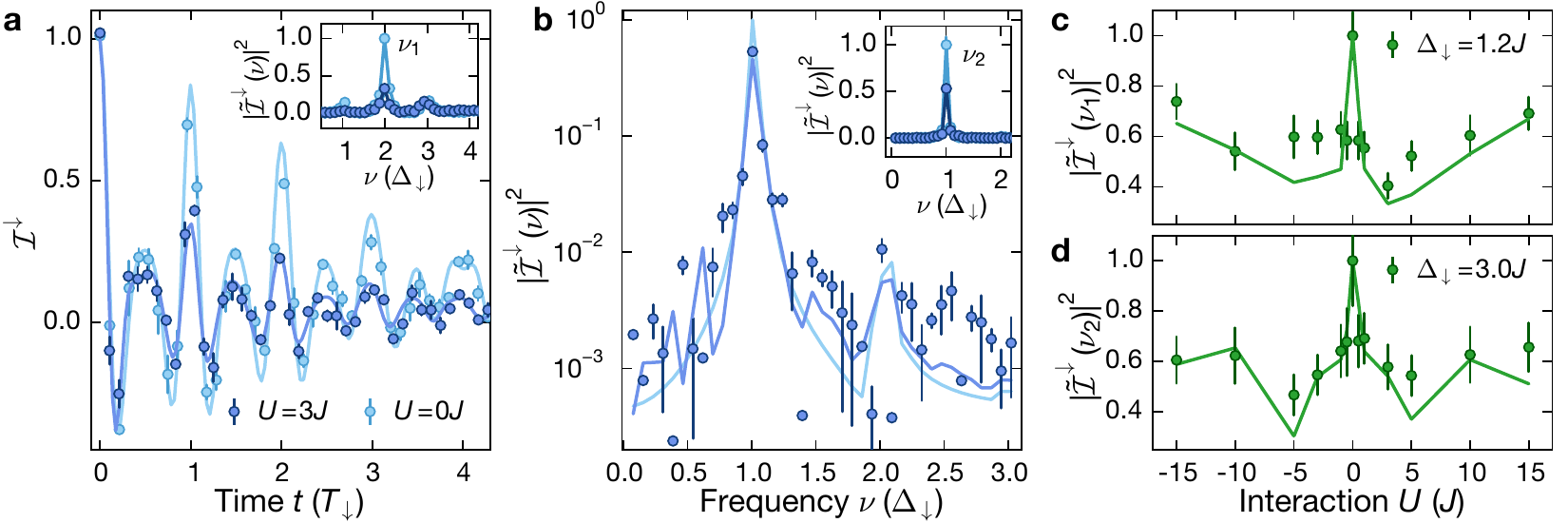}
	\caption{\textbf{Short-time interacting Bloch oscillations.} 
		\textbf{a} Imbalance $\mathcal{I}^{\downarrow}$ for $U=0J$ (spin-polarized gas, light blue) and $U=3J$ (spin-resolved measurement, dark blue) with $J/h=\SI{0.88(2)}{\kilo \hertz}$ and $\Delta_{\downarrow}=1.22(1)J$. Inset: Power spectral density (PSD) $|\tilde{\mathcal{I}}(\nu)|^2$ of the time traces shown in the main panel, normalized to the maximum of the non-interacting spectrum; $\nu_1=2 \Delta_{\downarrow}/h$ indicates the dominant frequency component. 
		\textbf{b} PSD $|\tilde{\mathcal{I}}(\nu)|^2$ for $U=3J$ (spin-resolved measurement, dark blue), normalized to the maximum of the non-interacting spectrum; $J/h=\SI{0.54(1)}{\kilo \hertz}$ and $\Delta_{\downarrow}=2.96(3)J$. The data was obtained from time-traces as in (a). Inset: PSD as in the main panel and for $U=0J$ (spin-polarized gas, light blue). $\nu_2= \Delta_{\downarrow}/h$ indicates the dominant frequency.
		\textbf{c,d} Interaction scan of the peak power spectral density $|\tilde{\mathcal{I}}(\nu_j)|^2$ evaluated by summing the PSD in a window of $\pm 3$ data points around the dominant frequency $\nu_{j}$, $j=\{1,2\}$ at (c)$\Delta_{\downarrow}=1.22(1)J$ and (d) $\Delta_{\downarrow}=2.96(3)J$ obtained from traces as in (a). Each data point in (a),(b) consists of four independent measurements and the error bars denote the standard error of the mean (SEM). Solid lines in all panels are numerical simulations using TEBD (Methods). }
	\label{fig2}
\end{figure*}

In a first set of measurements we study the effect of interactions on the coherent short-time dynamics.
In a tilted lattice an initially localized particle exhibits Bloch oscillations~\cite{ben_dahan_bloch_1996}, with a characteristic period $T_{\sigma}=h/\Delta_{\sigma}$, set by the spin-dependent tilt. In the presence of interactions, Bloch oscillations persist, showing a rich variety of dynamics, such as interaction-induced dephasing and amplitude modulation~\cite{buchleitner_interaction-induced_2003,kolovsky_floquet-bloch_2003,tomadin_many-body_2007,tomadin_many-body_2008,gustavsson_control_2008, preiss_strongly_2015}. 
Here, we use the spin-resolved imbalance to probe real-space Bloch oscillations in a parity-projected manner. 
In the non-interacting limit the time-dependence can be computed analytically:

\begin{equation}
\mathcal{I}^\sigma(t) =\mathcal{J}_{0} \left( \frac{8J}{\Delta_\sigma} \sin\left(\frac{\pi \Delta_\sigma t}{h}\right) \right),
\label{eq:Imbalance}
\end{equation} 

\noindent which enables a precise calibration of the model parameters $\Delta_\sigma$ and $J$ (Fig.~\ref{fig2}a) at short times. Here, $\mathcal{J}_{0}$ denotes the 0th-order Bessel function of the first kind. The dephasing of the oscillations is caused by a residual harmonic confinement that results in a weak local variation $\delta T_{\sigma}$ of the Bloch oscillation period $T_\sigma$ between adjacent sites. 
An upper bound for the trap frequency $\omega_h/(2 \pi)=\SI{39}{\hertz}$ was extracted from independent measurements~\cite{supplements} and corresponds to $\delta T_\sigma/T_\sigma \ll 10^{-3}$. Since the imbalance dynamics for both spin components is very similar (see Fig.~\ref{fig_spinres_BO}), we focus on one component $\mathcal{I}^\downarrow$.

For weak tilt values, $\Delta_{\downarrow}= 1.2J$, we find that the dynamics of the interacting spin-mixture ($U=3J$) exhibits the same dominant frequency components as the non-interacting Bloch oscillations, while the dephasing is strongly enhanced. This can be seen more directly by calculating the power spectral density (PSD) of the imbalance $|\tilde{\mathcal{I}}^\sigma(\nu)|^2$ (inset of Fig.~\ref{fig2}a). 
We find three distinct peaks in the spectrum, the Bloch frequency $\Delta_{\downarrow}$ and an admixture of two higher harmonics with the largest spectral weight in the second harmonic at $\nu_1=2\Delta_\downarrow/h$. For $U=3J$ its weight is decreased by $70 \%$ compared to the non-interacting case. The higher-order harmonics originate from the real-space evolution within one Bloch cycle and are determined by the  Bloch oscillation amplitude $A_{\sigma}/d_s=4J/\Delta_\sigma$. We anticipate frequency components at integer multiples of $\Delta_\sigma$, with an upper bound determined by $A_{\sigma}/d_s$, in agreement with our data. 

Interaction effects are expected to be less relevant once the Bloch oscillation amplitude is smaller than one site, resulting in negligible overlap between neighboring particles for our CDW initial state. In Fig.~\ref{fig2}b we show the PSD of the coherent short-time dynamics for $\Delta_{\downarrow}= 3.0J$. While the largest spectral weight of the PSD is now contained in the Bloch frequency $\nu_2=\Delta_\downarrow/h$, the reduction is still about $50 \%$ compared to the non-interacting case. Indeed, the spectral weight is a sensitive measure of the interaction-induced dephasing. Moreover, the on-site interactions lift the degeneracy of the energy levels in the Wannier-Stark spectrum, which results in additional frequency components in the PSD. For our parameters (Fig.~\ref{fig2}b) they occur at $\approx\nu_2 \pm 0.5 \Delta_\downarrow/h$ in the time-evolving block decimation (TEBD) simulations~\cite{schollwock_2016,paeckel_2019,hauschild_2018}, which is consistent with our data.

\begin{figure*}[t!]
	\includegraphics[width=\textwidth]{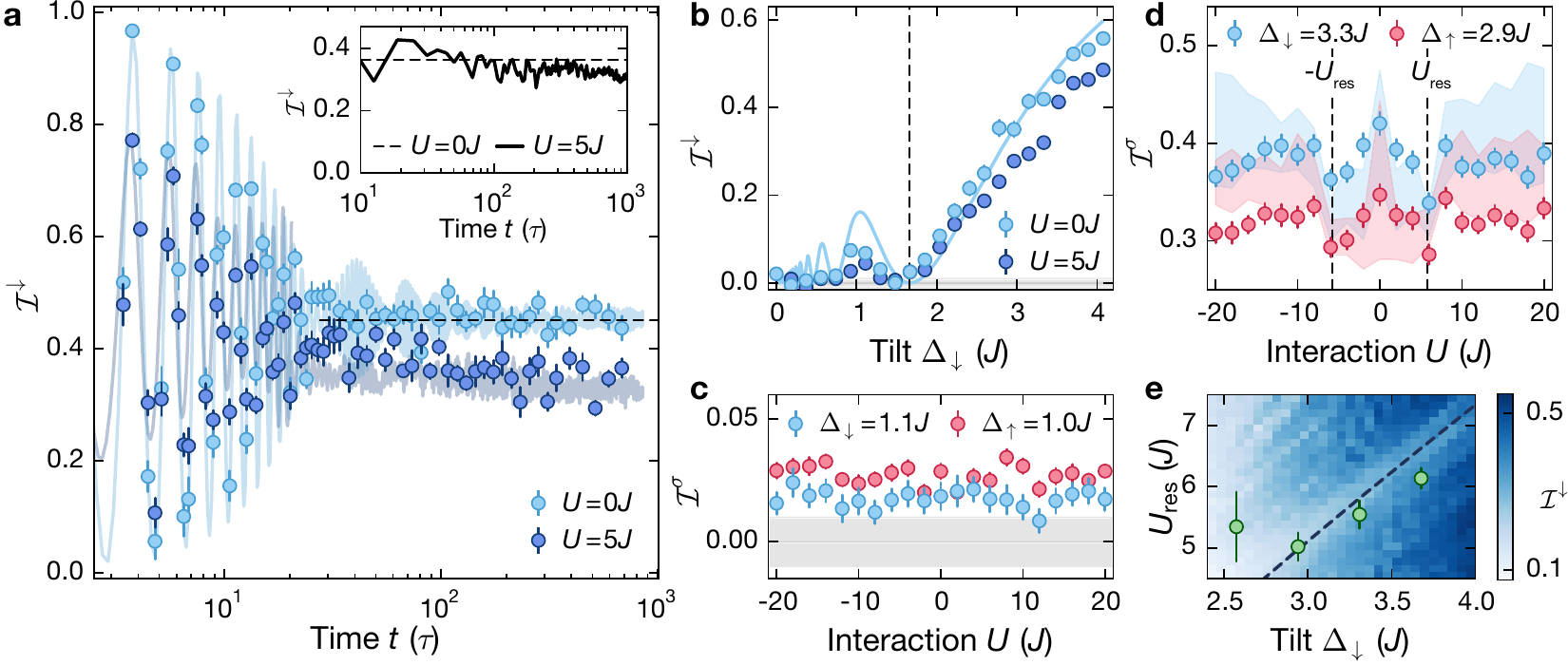}
	\caption{\textbf{Long-time dynamics.}
		\textbf{a} Imbalance time traces at $\Delta_{\downarrow}=3.30(3) J$ and $J/h= \SI{0.54(1)}{\kilo \hertz}$ for $U=0J$ (spin-polarized, light blue) and $U=5J$ (spin-resolved measurement, dark blue).  The shaded trace is an ED calculation for $L=16$ (Methods). Each data point is averaged over 12 individual experimental realizations. Inset: ED calculation for $L=16$ in a clean system with $\Delta_\downarrow=\Delta_\uparrow=3J$, $\omega_h=0$ and $U=5J$ using a N\'eel-ordered initial CDW. The dashed lines show the analytic prediction for the non-interacting steady-state imbalance [Eq.~\eqref{eq:Imbalance_state}].
		\textbf{b} Steady-state imbalance versus $\Delta_\downarrow$ measured at $U=0J$ (spin-polarized, light blue) and $U=5J$ (spin-resolved measurement, dark blue). Each data point is averaged over ten equally spaced times in a time window between $70 \tau$ and $100 \tau$ ($U=0J$) and  $340 \tau$ and $370 \tau$ ($U=5J$). The solid line shows the analytic prediction for $\mathcal{I}^\downarrow$ [Eq.~\eqref{eq:Imbalance_state}] and the dashed line indicates the first root of the Bessel function at $\Delta_{\downarrow}\approx1.5J$. 
		\textbf{c} Spin-resolved steady-state imbalance versus interaction strength at $\Delta_{\downarrow}=1.10(1) J$. Each point is averaged over ten time steps equally spaced between $170 \tau$ and $200 \tau$.  
		\textbf{d} Spin-resolved steady-state imbalance versus interaction strength as in (c) for $\Delta_{\downarrow}=3.30(3) J$. The shaded trace is an ED simulation, which is averaged over the same time steps as in (c) and where the width indicates the $1\sigma$ standard deviation. 
		\textbf{e} Resonances extracted from interaction scans for $U>0$ as in (d) for different tilt values~\cite{supplements}. The color plot shows ED calculations for the same parameters as in the experiment, but with $\omega_h=0$, for $L=13$ sites. The dashed line indicates the analytic prediction for the resonance $U_{\mathrm{res}} \simeq 2\Delta_\downarrow-8J^2/(3\Delta_\downarrow)$. The grey shaded area in (b),(c) indicates our calibrated detection resolution. In all panels error bars denote the SEM.}
	\label{fig3}
\end{figure*}

The sensitivity of the coherent short-time dynamics on the interaction strength is further highlighted by the strong interaction-dependence of the peak power spectral density (PPSD) $|\tilde{\mathcal{I}}(\nu_{j})|^2$ of the respective dominant frequency components $\nu_{j}$, $j=\{1,2\}$ (Fig.~\ref{fig2}c,d). We find a sharp decrease of the PPSD by about $40 \%$ already for small interaction strength $U= \pm 0.5J$ for $\Delta_\sigma=1.2J$. After reaching a global minimum at intermediate interaction strength, it slowly recovers to the non-interacting value in the limit of large interactions.


For long enough evolution times, the coherent Bloch oscillations are dephased and a finite steady-state imbalance develops in the non-interacting limit (Fig.~\ref{fig3}a). 
Note that, if the dephasing was solely due to residual harmonic confinement, we would expect a coherent revival of the oscillations, which is suppressed in our experiment by additional dephasing mechanisms and ensemble averaging. 
The observed finite steady-state imbalance is caused by Wannier-Stark localization and can be computed analytically by time averaging the short-time dynamics:

\begin{equation}
\mathcal{I}^\sigma= \lim_{T \to \infty} \frac{1}{T} \int_0^{T} \mathcal{I}^\sigma(t) \, dt =\mathcal{J}_0^2\left(\frac{4J}{\Delta_{\sigma}}\right).
\label{eq:Imbalance_state}
\end{equation} 

\noindent Excellent agreement between our data and the analytical result provides strong evidence that the effect of the harmonic confinement is negligible for the late-time steady-state imbalance, in contrast to previous  fermionic transport experiments~\cite{ott_collisionally_2004,strohmaier_interaction-controlled_2007}. This is further supported by the data in Fig.~\ref{fig3}b, where the steady-state value is probed for a larger range of tilt values, even reproducing the non-monotonous behavior that is found for small values of the tilt. Note, that the vanishing imbalance, as observed for $\Delta_\downarrow \approx 1.5J$ (dashed line in Fig.~\ref{fig3}b), does not indicate delocalization. It results from localized Wannier-Stark orbitals with equal weight on even and odd sites. 

In the presence of weak interactions  localization was predicted to survive in the limit of small additional disorder or harmonic confinement, signaled by a finite steady-state imbalance~\cite{schulz_stark_2019, nieuwenburg_bloch_2019}. Here, we find that after a small decay at intermediate times a plateau of the imbalance develops, which persists for long evolution times up to $700\tau$ (Fig.~\ref{fig3}a) in the strongly-interacting regime.  A comparison with ED simulations (inset Fig.~\ref{fig3}a) in a clean system without spin-dependent tilt and without harmonic confinement for a N\'eel-ordered initial CDW (as opposed to the random-spin initial state realized in the experiment) further highlights that this non-ergodic behavior is not due to experimental imperfections at least for the experimentally relevant observation times (see Fig.~\ref{fig:scaling} for a systematic finite-size scaling analysis). Moreover, this robust steady-state value survives over a wide range of parameters (Fig.~\ref{fig3}b). As a function of the tilt it qualitatively follows the behavior of the non-interacting system, but shows consistently lower steady-state values.

The persistence of non-ergodicity down to very small values of the tilt is surprising at first sight. One may expect that for large Bloch-oscillation amplitudes the interactions between particles result in a dephasing of the coherent dynamics that give rise to Wannier-Stark localization in the non-interacting limit and hence cause ergodic behavior~\cite{buchleitner_interaction-induced_2003,kolovsky_floquet-bloch_2003,tomadin_many-body_2007,tomadin_many-body_2008,ott_collisionally_2004,strohmaier_interaction-controlled_2007}. We study the plateau value for $\Delta_{\downarrow}=1.1 J$ and find that it is largely independent of interactions (Fig.~\ref{fig3}c). In a numerical analysis of this regime for a N\'eel-ordered singlon CDW we indeed find that the imbalance decays to zero for evolution times on the order of $10^4\,\tau$~(Fig.~\ref{fig:scaling}), which further agrees with the finite imbalance measured at $\sim 200\,\tau$. The observed inversion of the spin-resolved imbalance $\mathcal{I}^\downarrow < \mathcal{I}^\uparrow$ after long evolution times (although $\Delta_\downarrow > \Delta_\uparrow$) is explained by the non-monotonic dependence of the stationary imbalance on the tilt for $\Delta_{\sigma} <2J$ as shown in Fig.~\ref{fig3}b.

For intermediate values of the tilt $\Delta/J\simeq 3$ on the other hand, we find a surprisingly robust steady-state imbalance, in agreement with numerical calculations, with a clear interaction dependence (Fig.~\ref{fig3}d). The behavior is similar for both spin components and well reproduced by numerical simulations. The deviation between experiment and numerical simulations at larger interaction strengths is most likely due to the finite coupling between 1D chains, which plays a larger role for increased interactions~\cite{bordia_coupling_2016}.
The steady-state imbalance is symmetric around $U=0$ due to a dynamical symmetry $[$for $(\Delta_\downarrow-\Delta_\uparrow) \ll J ]$ between attractive and repulsive interactions~(Sect.~\ref{sec:Usymm}), similar to the homogeneous Fermi-Hubbard model~\cite{schneider_fermionic_2012,schreiber_observation_2015}.
The curve displays a global minimum for intermediate interactions, which we identify with resonant processes at $|U| \simeq 2\Delta$, where two singlons separated by two lattice sites form a doublon. This coincides with regime \ding{172} in Fig.~\ref{fig1}c, where the largest connectivities were found.
The precise value of the resonance is slightly shifted, $U_{\text{res}} \simeq 2\Delta-8J^2/(3\Delta)$, due to perturbative corrections for finite $J/\Delta$, 
in agreement with our data (dashed line in Fig.~\ref{fig3}e). 
For large interactions and weak spin-dependence $(\Delta_\downarrow-\Delta_\uparrow) \ll J$, we expect the system to recover the non-interacting regime~\cite{supplements}.

In order to gain additional insights into the observed non-ergodic behavior, we study the properties of our model perturbatively in the large tilt limit for the two distinct regimes \ding{172} and \ding{173} (Fig.~\ref{fig1}c). In regime \ding{173}, $\Delta\gg J, |U|$, an effective Hamiltonian can be derived in powers of $\lambda=J/\Delta$. As predicted~\cite{nieuwenburg_bloch_2019,moudgalya_thermalization_2019,sala_ergodicity_2020,khemani_localization_2020, taylor_experimental_2020}, we find an emergent dipole-conserving Hamiltonian $\hat{H}^{\text{dip}}_{\text{eff}}$ [Eq.~(\ref{eq:H3_dipole})] up to third order in $\lambda$~(Sect.~\ref{sect:effHam}), where the dipole-moment operator is defined as $\sum_{i, \sigma} i \hat n_{i, \sigma}$. The dominant off-diagonal terms of $\hat{H}^{\text{dip}}_{\text{eff}}$ are of similar nature as those in the fragmented Hamiltonians studied previously~\cite{sala_ergodicity_2020,khemani_localization_2020}, seemingly consistent with the observed non-ergodic behavior. Yet, higher-order processes $\mathcal{O}(\lambda^4)$, relevant for $\Delta \simeq 3J$, are expected to melt the CDW within the experimentally studied timescales~\cite{sala_ergodicity_2020}. These higher-order processes as well as the dominant off-diagonal contribution, however, require the production of doublons, which is penalized by the on-site interaction $U$. We  numerically show that this leads to a significant slowdown of the dynamics~(Sect.~\ref{sect:effHam}), which explains the robustness of the steady-state value observed in the experiment. Thus, for large values of the tilt, the doublon number is effectively conserved as well, as suggested in Ref.~\cite{nieuwenburg_bloch_2019}.

\begin{figure}[t!]
	\includegraphics[width=3.3in]{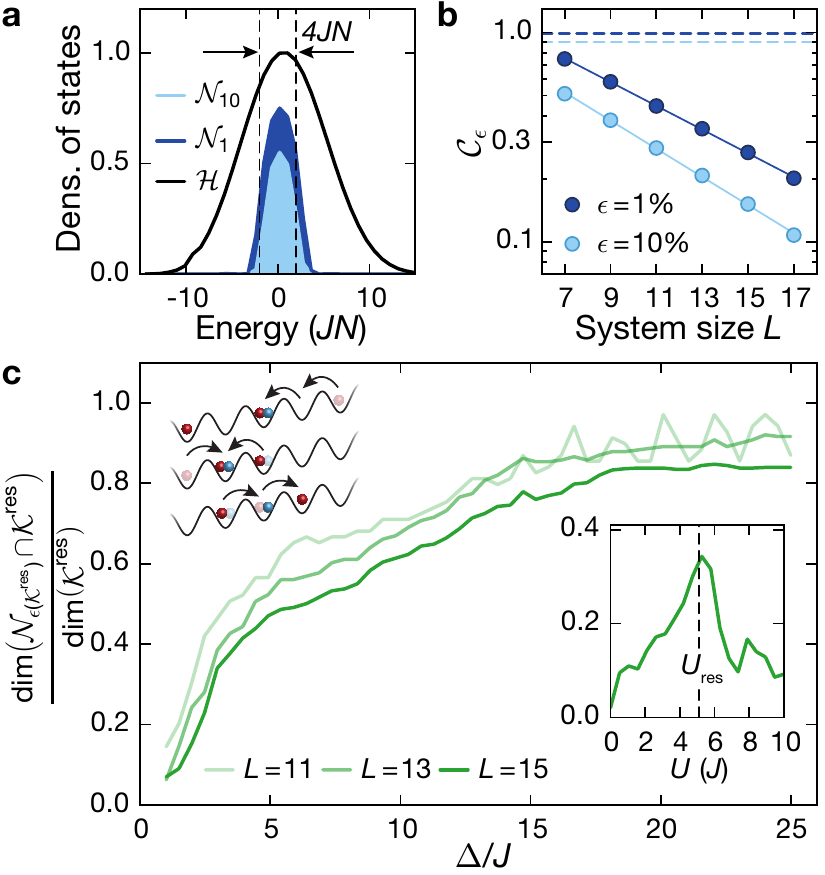}	 
	\caption{\textbf{Theoretical analysis of the relevant many-body states for $\omega_h=0$, $\Delta_\uparrow=\Delta_\downarrow\equiv\Delta$ and a N\'eel-ordered initial state.} 
		\textbf{a} Density of states in the full Hilbert space $\mathcal{H}$ restricted to quarter filling and zero magnetization for the numerical fragments $\mathcal{N}_{1}$ ($\epsilon=1\%$), $\mathcal{N}_{10}$ ($\epsilon=10\%$), $U=5J$, $\Delta=3J$ and $T_\mathcal{N}=1000\tau$, normalized to the maximum in $\mathcal{H}$; $L=15$. 
		\textbf{b} Scaling of the finite-time connectivity $\mathcal{C}_\epsilon$ with system size for a time window $T_\mathcal{N}=1000 \tau$, $U=5J$ and $\Delta=3J$. Solid lines are exponential fits to the data. Dashed lines are the prediction for the finite-time connectivity of a thermal state, showing a constant scaling at $1-\epsilon$. 
		\textbf{c} Normalized intersection for $U=U_{\text{res}}$ between the Krylov subspace $\mathcal{K^{\text{res}}}$ and the numerical fragment $\mathcal{N}_{\epsilon(\mathcal{K^{\text{res}}})}$ (Methods), where $\text{dim}(\mathcal{N}_{\epsilon(\mathcal{K^{\text{res}}})})=\text{dim}(\mathcal{K^{\text{res}}})$ (main text). The schematic shows the most important processes, connecting the states within the Krylov subspace $\mathcal{K^{\text{res}}}$~\cite{supplements}. Inset: Normalized intersection as in the main plot for $\Delta=3J$. The dashed line illustrates the resonance condition found in regime \ding{172}.}
	\label{fig4}
\end{figure}


On resonance, $|U| \simeq 2 \Delta$ (regime \ding{172} in Fig.~\ref{fig1}c), doublons can be formed without energy penalties, possibly leading to faster dynamics. Indeed, after an initial faster dynamics, we find a lower steady-state imbalance, which cannot be solely explained by the second-order resonant tunneling process shown in Fig.~\ref{fig1}c, because it leaves the imbalance invariant. 
In this regime, we derive an effective Hamiltonian $\hat{H}^{\text{res}}_{\text{eff}}$~[Eq.~(\ref{eq:H2_resonance})] up to second order in $\lambda$ (the third order vanishes), conserving the dipole moment, the doublon number or the sum of the two ($\sum_{i, \sigma} i \hat n_{i, \sigma} + 2\sum_i \hat n_{i,\uparrow} \hat n_{i,\downarrow}$). 
The corresponding symmetry sector exhibits strong fragmentation and results in a finite steady-state imbalance~\cite{supplements}.
In Fig.~\ref{fig4}c we show the dominant second-order tunneling terms for our initial state, illustrating the importance of doublon-assisted tunneling processes for the reduction of the steady-state imbalance. 
For finite $\lambda$ or longer evolution times, higher-order hopping processes $\mathcal{O}(\lambda^4)$ enable additional dynamics. These processes are expected to eventually melt the CDW completely, although the required timescales may be very large. In the experiment, we find robust steady-state values even for rather low values of the tilt ($\Delta\simeq 3J$) up to evolution times of about 700$\tau$ (Fig.~\ref{fig3}a).

In order to connect the large-tilt limit described by $\hat{H}^{\text{res}}_{\text{eff}}$ to the experimental parameter regime, we investigate the states within the explored subspace $\mathcal{N}_\epsilon$, which we denote numerical fragment in analogy to the phenomenon of Hilbert-space fragmentation. For simplicity, we study a clean system ($\omega_h=0$, $\Delta_\uparrow=\Delta_\downarrow\equiv\Delta$) and a N\'eel-ordered CDW initial state. In Fig.~\ref{fig4}a, we show the density of states in the Hilbert space $\mathcal{H}$ and compare it to the density of states in the numerical fragment $\mathcal{N}_\epsilon$ for different values of the cut-off $\epsilon$. Centered around the energy of the initial state, the density of states acquires a finite width within the numerical fragments, that is approximately set by the many-body bandwidth $\pm 2 J N$ (dashed line in Fig.~\ref{fig4}a), where $N=N^\uparrow+N^\downarrow$ denotes the total number of atoms. 
In stark contrast to thermal systems, the low finite-time connectivity indicates that only a small number of states is relevant for the dynamics.
Moreover, it vanishes exponentially in the thermodynamic limit for finite evolution times up to $1000\,\tau$ (Fig.~\ref{fig4}b).   
Since the perturbative Hamiltonian $\hat{H}^{\text{res}}_{\text{eff}}$ is only valid in the limit of large tilts, the intersection between the numerically constructed fragment and the analytical one $\mathcal{K}^{\text{res}}$  (Methods), which was derived using the perturbative Hamiltonian $\hat{H}_\text{eff}^{\text{res}}$ up to third order in $\lambda=J/\Delta$, is small for our experimental parameters $\Delta=3J$ and $U=5J$ (Fig.~\ref{fig4}c). 
We expect, however, that the two subsectors coincide for $\lambda\rightarrow 0$. Indeed the normalized intersection saturates to one, although only for $\Delta/J \gg 20$. For this comparison the cut-off value $\epsilon ( \mathcal{K^{\text{res}}} ) $ is chosen such that $\text{dim}(\mathcal{N}_{\epsilon ( \mathcal{K^{\text{res}}} ) })=\text{dim}(\mathcal{K^{\text{res}}})$, since generally, $\mathcal{N}_\epsilon$ contains a much larger number of states. Despite the large value of $\lambda$ realized in the experiment, we find strong evidence that the slow dynamics is due to kinetic constraints and that the energetically allowed microscopic processes give rise to the phenomenon of Hilbert-space fragmentation in the large tilt limit, as demonstrated for the two regimes (\ding{172} and \ding{173}). This is further supported by the resonance feature that is shown in the inset of Fig.~\ref{fig4}c for the resonant regime \ding{172}.


In conclusion, we have demonstrated both experimentally and numerically non-ergodic behavior in the tilted 1D Fermi-Hubbard model over a wide range of parameters and have provided a microscopic understanding based on perturbative analytical calculations. 
For future studies it would be interesting to study the limit of large tilts, where strongly-fragmented effective Hamiltonians were identified and to investigate the initial-state dependence of the transient dynamics. This is a characteristic feature of Hilbert-space fragmentation, where distinct thermalization properties are expected for different fragments~\cite{moudgalya_thermalization_2019,khemani_localization_2020,sala_ergodicity_2020}. Although experimentally challenging due to finite evolution times, it would be interesting to reconcile the phenomenon of Stark MBL and Hilbert-space fragmentation, by studying the impact of weak disorder or residual harmonic confinement on the long-time dynamics~\cite{taylor_experimental_2020}.
Adding periodic modulation as an additional ingredient, other strongly-fragmented models, scarred models and time crystals could be engineered~\cite{pai_dynamical_2019,zhao_quantum_2020,kshetrimayum_stark_2020} or drive-induced localization could be investigated~\cite{bairey_driving_2017,bhakuni_drive-induced_2020}. By tuning the direction of the tilt in a 2D lattice, dipole- and higher-moment conserving models could be realized~\cite{khemani_localization_2020,feldmeier_anomalous_2020} enabling studies beyond the hydrodynamic regime~\cite{guardado-sanchez_subdiffusion_2020}.
Moreover, it will be interesting to explore the connection between lattice gauge theories and the phenomenon of Hilbert-space fragmentation~\cite{smith_disorder-free_2017,brenes_many-body_2018,pai_fractons_2020,verdel_real-time_2020,rakovszky_statistical_2020,yang_hilbert-space_2020}, which could be addressed experimentally in a similar model~\cite{yang_observation_2020}.


\paragraph*{\textbf{Acknowledgements}}
We thank D.~Abanin,  G.~De Tomasi, M.~Filippone, M.~Knap, N.~Lindner, R.~Moessner, T.~Rakovszky and N.~Yao for inspiring discussions. We thank C.~Schweizer for very useful discussion about the experimental results and their interpretation. We thank M.~Buser for illuminating discussions regarding the ED calculations. This work was supported by the Deutsche Forschungsgemeinschaft (DFG, German Research Foundation) under Germany's Excellence Strategy -- EXC-2111 -- 39081486. The work at LMU was additionally supported by DIP and B.~H.~M. acknowledges support from the European Union (Marie Curie, Pasquans). The work at TU was additionally supported by the European Research Council (ERC) under the European Union’s Horizon 2020 research and innovation program (grant agreement No. 771537).

\paragraph*{\textbf{Author contributions}}
S.S., T.K. and B.H.M. conceived and performed the experiments and analyzed the data. B.H.M. and P.S. carried out the numerical simulations. P.S. carried out the analytic derivations. M.A., F.P. and I.B. supervised the work. All authors contributed critically to the writing of the manuscript and the interpretation of experimental and numerical results.

\paragraph*{\textbf{Data availability}}
The data that support the plots within this paper and other findings of this study are available from the corresponding author upon reasonable request.

\paragraph*{\textbf{Code availability}}
The code that supports the plots within this paper are available from the corresponding author upon reasonable request.

\paragraph*{\textbf{Competing interests}}
The authors declare no competing interests.

\addtocontents{toc}{\string\tocdepth@munge}
\section*{Methods}
\addtocontents{toc}{\string\tocdepth@restore}

\paragraph*{\textbf{Experimental sequence.}}

Our sequence begins with loading a degenerate Fermi gas with temperature $T/T_F=0.15(1)$, where $T_F$ is the Fermi temperature, into a three-dimensional (3D) optical lattice. The wavelength is  $\lambda_l = \SI{1064}{nm}$ along the $x$ direction and $\lambda_\perp =  \SI{738}{nm}$ in the transverse directions. Repulsive interactions during loading in combination with a short, off-resonant light pulse after loading ensure an initial state free of double occupancies~\cite{supplements}. 
By adding a short lattice with wavelength $\lambda_s=\lambda_l/2$ along the $x$ direction, we generate a CDW initial state consisting of singlons~\cite{supplements}. Holding the gas in this deep 3D lattice with a tilted, bichromatic superlattice along the $x$ direction, dephases remaining correlations between neighboring sites and suppresses any residual dynamics, while ramping up a magnetic field gradient and adjusting the interaction strength. 
The lattice depths are $18 \, E_{rs}$ for the short lattice, $20 \, E_{rl}$ for the long lattice and $55 \, E_{r\perp}$ for the transverse lattices. The depths are given in the respective recoil energies, $E_{rj}=\hbar^2 k_{j}^2/(2m)$, with $j\in\{l,s,\perp \}$, $k_{j}= 2 \pi/\lambda_{j}$ the corresponding wave vector, $m$ the mass of $^{40}\mathrm{K}$ and and $\hbar=h/(2\pi)$ the reduced Planck constant. 
The deep transverse lattices decouple the 1D chains aligned along $x$ and generate a 2D array of nearly independent 1D systems. The residual coupling along the transverse directions is typically less then 0.03 $\%$ of the coupling $J$ along $x$.
The dynamics to probe the  tilted 1D Fermi-Hubbard model described by the Hamiltonian in Eq.~\eqref{eq:Hamiltonian} is initiated by suddenly switching off the long lattice and quenching the short lattice to depths between $6 \, E_{rs}$ and $ 8 \, E_{rs}$. Simultaneously, the strength of the dipole trap is adjusted in order to compensate the anti-confining harmonic potential introduced by the lattice~\cite{supplements}. After a variable evolution time $t$ the on-site population is frozen by suddenly ramping up the longitudinal lattices to $18 \, E_{rs}$ and $20 \, E_{rl}$ respectively. Subsequently, we extract the spin-resolved imbalance $\mathcal{I}^{\sigma}$, by using a bandmapping technique~\cite{sebby-strabley_lattice_2006,foelling_direct_2007} in conjunction with Stern-Gerlach resolved absorption imaging.

\paragraph*{\textbf{Initial state.}}
The initial state in all experiments consists of a CDW of singlons, where $\ket {\uparrow}$ and $\ket {\downarrow}$ states are randomly distributed on even lattice sites and odd lattice sites are empty. We work with an equal mixture of both states ($N_{\uparrow}=N_{\downarrow}$) such that the total magnetization is zero. The fraction of residual holes on even lattice sites, due to imperfections in the loading sequence and due to removed doublons is expected to be about $10\%$~\cite{scherg_nonequilibrium_2018}. Excellent agreement between the data and numerical simulations, which do not consider residual holes on even sites, indicates, that the hole fraction has a  negligible effect on our dynamics.
The initial state can be modelled as incoherent mixture within the zero magnetization sector with density matrix $\hat \rho=\frac{1}{\mathcal{N}}\sum_{\{\sigma\}|\sum_i \sigma_i=0}  \ket{\psi_0(\{\sigma\})}\bra{\psi_0(\{\sigma\})}$, where each product state  $\ket{\psi_0(\{\sigma\})}$, is given by a CDW of singlons and where the sum runs over all $\mathcal{N}$ possible permutations of spin configurations $\{\sigma\}$. The product state $\ket{\psi_0(\{\sigma\})}$ is defined as $\ket{\psi_0(\{\sigma\})} = \prod_{i=\text{even}\in {\rm trap}} \left(\hat{c}_{i\uparrow}^{\dagger}\right)^{n_{i\uparrow}} \left(\hat{c}_{i\downarrow}^{\dagger}\right)^{n_{i\downarrow}} \ket{0}$, where $\hat{c}_{i \sigma}^{\dagger}$ is the fermionic creation operator, $n_{i\sigma} \in \{0,1\}$, $\sigma \in \{ \uparrow, \downarrow \} $, $n_{i}=n_{i\uparrow}+n_{i\downarrow}\leq 1 $ and $i$ is the lattice-site index along $x$.

\paragraph*{\textbf{Details of numerical calculations}}

The numerical computations that are compared with the experiment in Fig.~\ref{fig2} and Fig.~\ref{fig3} of the main text were performed using ED or TEBD. The parameters $J$, $\Delta_{\uparrow}$ and $\Delta_{\downarrow}$ used in the computations were obtained as fit parameters from the corresponding non-interacting data. Additionally, the effect of harmonic confinement present in the experiment was simulated by scaling the trap frequency by a factor $\sqrt{\frac{L_\text{exp}}{L}}$ where $L_\text{exp}=290$ is the system size in the experiment and $L$ is the system size used in the numerical calculation. This is done so as to appropriately simulate the collapse and revival dynamics in the Bloch oscillations induced by the harmonic confinement~\cite{supplements}. 

We use TEBD for short-time dynamics (Fig.~\ref{fig2} of the main text) and ED for long-time dynamics (Fig.~\ref{fig3} of the main text). In ED, we consider the Hilbert space as a tensor product $\mathcal{H}_{\uparrow}\otimes \mathcal{H}_{\downarrow}$ where $\mathcal{H}_{\sigma}$ is the Hilbert space of spin-$\sigma$ atoms. In order to efficiently compute the time dynamics, we decompose each time step in the dynamics into three unitary propagators. One each corresponding to the hopping of the two spin components and the third one corresponding to the on-site potential and interactions. We use a Trotter-Suzuki approximation in this decomposition (see Supplementary Information~\cite{supplements} for details and error analysis). 
In Fig.~\ref{fig3}a,d, we use $L=16, N_{\uparrow}=N_{\downarrow}=4$. In order to effectively model a mixed CDW initial state, in Fig.~\ref{fig3}a, this computation is averaged over $20$ randomly chosen pure CDW states. In Fig.~\ref{fig3}d we use a superposition of pure CDW product states as we are concerned only with time-averaged steady-state value. The parameters $J$, $\Delta_{\sigma}$ and the harmonic confinement are fixed by fitting to the corresponding non-interacting data. 

In Fig.~\ref{fig2}, we use TEBD calculations with $L=100$ and bond-dimension $\chi=120$. The truncation error  was less than $10^{-2}$. In Fig.~\ref{fig2}b,c, we compare the experimental and numerical data in Fourier space. If the two data sets have different number of samplings in the time domain, we scale the numerical data appropriately after the fast Fourier transform.

\paragraph*{\textbf{Construction of the Krylov subspace}}

The Krylov subspace (corresponding to the fragment $\mathcal{K^{\text{res}}}$) is constructed by using the effective Hamiltonian on resonance $\hat H^{\mathrm{res}}_{\mathrm{eff}}$ in Eq.~\eqref{eq:H2_resonance}. This Hamiltonian is then interpreted as an adjacency matrix in the Wannier basis and the Krylov subspace consists of all states, which are connected to the N\'eel-ordered CDW initial state. The Krylov subspace $\mathcal{K^{\text{res}}}$ is closed under time-evolution generated by the effective Hamiltonian $\hat H^{\mathrm{res}}_{\mathrm{eff}}$. Starting from initial states within the Krylov subspace $\mathcal{K^{\text{res}}}$ and including higher-order terms $\mathcal{O}(\lambda^4)$, the dynamics is captured only approximately~\cite{supplements}. An improvement is obtained by further rotating the diagonal basis in which the effective Hamiltonian becomes fragmented with the unitary transformation obtained in powers of $\lambda$ (as given by the Schrieffer-Wolff perturbative expansion~\cite{supplements}). This results in a rotated Krylov subspace.

\paragraph*{\textbf{Construction of the numerical fragment}}

We define the numerical fragment $\mathcal{N}_\epsilon$ as the span of a subset $\mathcal{B}_{\epsilon}$ of the number basis $\mathcal{B}$ of $\mathcal{H}$, where $\mathcal{H}$ is restricted to quarter filling and zero magnetization. We define the set $\mathcal{B}_{\epsilon}$ via its complement,  $\mathcal{B}_\epsilon=\mathcal{B} \backslash \mathcal{B}_\epsilon^c$, where $\mathcal{B}_{\epsilon}^c$ would be ideally defined as the largest subset of $\mathcal{B}$ satisfying $ \mathrm{max}_{t<T_{\mathcal{N}}} \sum_{n^c\in \mathcal{B}_{\epsilon}^c}  \left| \braket{n^c  | \psi(t)}\right|^2 < \epsilon$.  Here $T_{\mathcal{N}}$ defines a time window for the evolution of the initial state $\ket{\psi(t=0)}$. Equivalently, one could define the subset $\mathcal{B}_{\epsilon}$ as the smallest one, satisfying $\mathrm{min}_{t<T_{\mathcal{N}}} \sum_{n\in \mathcal{B}_{\epsilon}}  \left| \braket{n  | \psi(t)}\right|^2 \geq 1-\epsilon$. We work with the complement, because it is easier to implement numerically. This inequality condition for the complement would ensure that the residual overlap of $\ket{\psi(t)}$ outside of $\mathcal{N}_\epsilon$ at any time $t\leq T_{\mathcal{N}}$ is bounded by $\epsilon$. Constructing this $\mathcal{B}_{\epsilon}^c$, however, involves a search in the powerset of $\mathcal{B}$, which is exponential in the dimension of $\mathcal{H}$. This is intractable even for relatively small system sizes such as $L=7$. It follows from the inequality $  \mathrm{max}_{t<T_{\mathcal{N}}} \sum_{n^c}  \left| \braket{n^c  | \psi(t)}\right|^2 \leq  \sum_{n^c} \mathrm{max}_{t<T_{\mathcal{N}}} \left| \braket{n^c  | \psi(t)}\right|^2$ that keeping the latter sum smaller than $\epsilon$ will ensure that the former sum is also bounded by $\epsilon$. Moreover, the latter sum is computationally easier to handle and therefore, we use it to define the fragment. We construct the numerical fragment $\mathcal{N}_\epsilon$ using a $ \mathcal{B}_\epsilon^c$, defined such that $\sum_{n^c} \mathrm{max}_{t<T_{\mathcal{N}}} \left| \braket{n^c  | \psi(t)}\right|^2 < \epsilon$. The gap in the inequality $  \mathrm{max}_{t<T_{\mathcal{N}}} \sum_{n^c}  \left| \braket{n^c  | \psi(t)}\right|^2 \leq  \sum_{n^c} \mathrm{max}_{t<T_{\mathcal{N}}} \left| \braket{n^c  | \psi(t)}\right|^2$ loosely depends on the sum $ \sum_{n\in \mathcal{B}} \mathrm{max}_{t<T_{\mathcal{N}}} \left| \braket{n  | \psi(t)}\right|^2$, which is in general, not normalized. Although this sum can be as large as the dimension of $\mathcal{H}$, in the examples that we study, it remains small, i.e., $<10$ for $L<20$, and grows logarithmically in the dimension of $\mathcal{H}$.

\cleardoublepage


\setcounter{figure}{0}
\setcounter{equation}{0}
\setcounter{page}{1}

\newcommand{\downuparrows}{\mathbin\downarrow\hspace{-.5em}\uparrow}
\newcommand{\avg}[1]{\left< #1 \right>}

\renewcommand{\thepage}{S\arabic{page}} 
\renewcommand{\thesection}{S\arabic{section}} 
\renewcommand{\thetable}{S\arabic{table}}  
\renewcommand{\thefigure}{S\arabic{figure}} 
\renewcommand{\theequation}{S\arabic{equation}} 
\newcommand{\up}{\uparrow}
\newcommand{\dn}{\downarrow}

\renewcommand{\thesection}{}
\renewcommand{\thesubsection}{S\arabic{subsection}}

\def\tocname{Table of contents}

\onecolumngrid

\addtocontents{toc}{\string\tocdepth@munge}

\section*{\Large{supplementary information}}

\addtocontents{toc}{\string\tocdepth@restore}

\twocolumngrid

\setlength{\intextsep}{0.8cm} 
\setlength{\textfloatsep}{0.8cm}

In this supporting material we give a detailed overview of the analytical derivations, the experimental setup, the measurement techniques, the data acquisition and numerical techniques, employed in this work.

\tableofcontents

\subsection{Interaction picture}

The Hamiltonian~\eqref{eq:Hamiltonian} does not commute with spatial translations because of the tilted field. However, due to the gauge covariance of the Schrödinger equation, we can transform from the Schrödinger picture  to the interaction picture, where the Hamiltonian becomes translational invariant and time-periodic~\cite{murillo_manifold_2014, murillo_two-band_2013}.
%
We apply a unitary transformation $\hat{T}(t)\equiv e^{it \hat{H}_0 } $ with $\hat{H}_0 =  \sum_{\sigma}\Delta_{\sigma}\sum_i i \hat{n}_{i,\sigma}$ according to (we set $\hbar=1$ in the following)
\begin{equation}
\hat{H}_I(t)=\hat{T}(t)\hat{H} \hat{T}^{\dagger}(t) - i \hat{T}(t)\partial_t \hat{T}^{\dagger}(t),
\label{eq:Hint}
\end{equation}

\noindent with

\begin{align} \label{eq:HI}
\hat{H}_I(t)=&-J\sum_{i,\sigma}\big( e^{-i\Delta_{\sigma} t}{\hat{c}}_{i,\sigma}^{\dagger}{\hat{c}}_{i+1,\sigma} + \textrm{h.c}.\big) \\ \nonumber
&+ U\sum_i \hat{n}_{i,\uparrow}\hat{n}_{i,\downarrow},
\end{align}

\noindent which for incommensurate $\Delta_{\uparrow}\neq \Delta_{\downarrow}$ gives rise to a quasi-periodic Hamiltonian~\cite{else_long-lived_2020}.
Since density operators are gauge invariant
$\hat n_{i,I}(t)=\hat T(t) \hat n_{i} \hat T^{\dagger}(t)= \hat n_{i}$,
we can study the evolution and long-time value of the imbalance in the interaction picture, starting from the experimentally prepared charge-density wave (CDW) configuration without requiring an additional change of frame.
We emphasize that the Hamiltonian in the interaction picture [Eq.~\eqref{eq:HI}] explicitly commutes with lattice translations, is well-defined in the thermodynamic limit \cite{murillo_two-band_2013,nieuwenburg_bloch_2019} and avoids the superextensive-scaling contribution of the potential energy to the total energy of the system. 

For $\Delta_\uparrow=\Delta_\downarrow\equiv\Delta$, the interacting Hamiltonian [Eq.~\eqref{eq:HI}] can be understood as a periodically-driven system~\cite{goldman_periodically_2014,eckardt_high-frequency_2015,bukov_universal_2015}. 
For generic ergodic Hamiltonians, we expect the system to heat up to infinite temperature in the intermediate driving-frequency regime, which we probe in the experiment as a consequence of the energy absorption from the external drive~\cite{abanin_exponentially_2015, dalessio_long-time_2014}.
Therefore, we anticipate the density distribution to become homogeneous and the imbalance to decay to zero at infinite times. 
In order to give a lower bound on this heating timescale, we can make use of the rigorous theory of prethermalization for periodic \cite{abanin_exponentially_2015,abanin_effective_2017, de_roeck_very_2019, else_prethermal_2017, kuwahara_floquet-magnus_2016} or quasi-periodic Hamiltonians (for $\Delta_{\uparrow}\neq \Delta_{\downarrow}$) \cite{else_long-lived_2020} in the large tilt $\Delta\gg J$ regime~\footnote{This discussion would also apply in the limit of incommensurate $\Delta_{\sigma}$ and $U$ when going to the frame with respect to the interacting term.}. 
This predicts a heating timescale $\tau_*\sim \tau e^{c \Delta/\mu}$ for the former, where $c$ is a numerical constant of order one and $\mu\sim (J+U)$. 
However, for the tilts employed in the experiment $\Delta\sim 3J$, this lower bound is far from the experimentally observed non-ergodicity until times $t\sim 1000 \tau$.

\subsection{Dynamical symmetry: $U\to -U$}
\label{sec:Usymm}

According to the theorem proven in Ref.~\cite{schneider_fermionic_2012} (Supplementary material section SD), the Fermi-Hubbard model exhibits a dynamical symmetry between repulsive and attractive interactions for any observable, which is invariant under both time-reversal and $\pi$-boost $\hat{B}_Q=e^{i\pi \sum_{i,\sigma} i \hat{n}_{i, \sigma}}$, when considering initial states, that are time reversal invariant and only acquire a global phase under the $\pi$-boost transformation. While our interaction scans  in the main text are consistent with this symmetry, the assumptions are not valid in the presence of a tilt. We can, however, generalize these assumptions and show that the dynamical symmetry holds for our system as well.
Under a spatial inversion  $\hat{\mathcal{P}}$, i.e. sending $i \to -i$ with respect to the center of a finite chain with length $L$, the tilted potential of the Hamiltonian changes sign 

\begin{equation} \label{eq:H1}
\hat{H}(U,\Delta_{\uparrow},\Delta_{\downarrow}) \overset{\hat{\mathcal{P}}}{\to} \hat{H}(U,-\Delta_{\uparrow},-\Delta_{\downarrow}) \, . 
\end{equation}

\noindent Using the $\pi$-boost $\hat{B}_Q$ together with the inversion $\hat{\mathcal{P}}$

\begin{equation}
\hat{\mathcal{P}}\hat{B}_Q\hat{H}(U,\Delta_{\uparrow},\Delta_{\downarrow})\hat{B}_Q^\dagger\hat{\mathcal{P}}^{\dagger}=
-\hat{H}(-U,\Delta_{\uparrow},\Delta_{\downarrow})
\end{equation}

\noindent an equation similar to Eq.~(S11) in~\cite{schneider_fermionic_2012} can be obtained.
The experimental observable is the spin-resolved imbalance $\mathcal{\hat I}^\sigma=\sum_{i=-\frac{L}{2}}^{\frac{L}{2}}(-1)^i \hat{n}_{i,\sigma}$, which is invariant under inversion $\mathcal{\hat I}^\sigma\overset{\hat{\mathcal{P}}}{\to}\mathcal{\hat I}^\sigma$ and $\pi$-boost $\mathcal{\hat I}^\sigma\overset{\hat{B}_Q}{\to}\mathcal{\hat I}^\sigma$, but breaks time-reversal symmetry $\hat{\mathcal{T}}$. This symmetry is violated, because the spin degrees of freedom of the density operator $\hat{n}_{i,\sigma}$ are exchanged.

Assuming that $\Delta_{\uparrow}=\Delta_{\downarrow}$, the Hamiltonian has an additional SU$(2)$ spin symmetry and is invariant under spin-rotations around $\hat S^x=\sum_{\beta, \gamma=\uparrow, \downarrow} 1/2 \hat c^\dagger_\beta \sigma^x_{\beta \gamma} \hat c_\gamma$, where $\sigma^x_{\beta \gamma}$ are the matrix elements of the Pauli matrix. The local observable $\hat{n}_{i,\sigma}$ is invariant under the product of time reversal $\hat{\mathcal{T}}$ and $\pi$-rotations around $x$, and thus we obtain for the time-evolved imbalance operator $\hat{\mathcal{I}}^\sigma_{(U,\Delta_{\uparrow},\Delta_{\downarrow})}$   

\begin{equation}
\begin{split}
\hat{\mathcal{P}} \hat B_Q e^{-i \pi \hat S^x } \hat{\mathcal{T}} &\hat{\mathcal{I}}^\sigma_{(U,\Delta_{\uparrow},\Delta_{\downarrow})}(t) \hat{\mathcal{T}}^{-1} e^{i \pi \hat S^x } \hat B^\dagger_Q \hat{\mathcal{P}}^\dagger   = \\ 
&\hat{\mathcal{I}}^\sigma_{(-U, \Delta_{\uparrow},\Delta_{\downarrow})}(t).
\end{split}
\end{equation}

\noindent As long as $\Delta_{\downarrow}-\Delta_{\uparrow}$ is sufficiently small, an approximate dynamical symmetry is present for our observable.

We next focus on the required symmetries of the initial state. For all experiments, 
we consider initial states that are an incoherent sum within the zero magnetization sector (thus $N_{\uparrow}=N_{\downarrow}$) with density matrix $\hat \rho=\frac{1}{\mathcal{N}}\sum_{\{\sigma\}|\sum_i \sigma_i=0}  \ket{\psi_0(\{\sigma\})}\bra{\psi_0(\{\sigma\})}$, where each product state  $\ket{\psi_0(\{\sigma\})}$, is given by a CDW of singlons.
The sum runs over all possible permutations  $\{\sigma\}$ of the spins within the zero magnetization sector. Under the combined action of time reversal and $\pi$-rotation around $x$, this state is left invariant up to a global phase. This is also the case for the $\pi$-boost $\hat B_Q$. Moreover under spatial inversion $\hat{\mathcal{P}}$ a configuration $\{\sigma_i\}$ is mapped onto another one $\{\sigma_i^{\prime}\}$ appearing in the mixed state $\hat \rho$ with equal weight. Thus, the mixed state is also invariant under $\hat{\mathcal{P}}$. In conclusion, we find for our initial states

\begin{equation}
\mathcal{I}^\sigma_{(U,\Delta_{\uparrow},\Delta_{\downarrow})}(t)=\mathcal{I}^\sigma_{(-U, \Delta_{\uparrow},\Delta_{\downarrow})}(t).
\end{equation}

\noindent Note that this dynamical symmetry is weakly broken by experimental imperfections such as the harmonic confinement (see Sec.~\ref{sec:hom_pot}) and varying onsite-interaction strength (see Sec.~\ref{sec:int_averaging}).

\subsection{Effective Hamiltonians}
\label{sect:effHam}

In this section we analytically derive several effective Hamiltonians, starting from the clean tilted Fermi-Hubbard model  without harmonic confinement and without spin-dependent tilt, described by the Hamiltonian

\begin{align} 
\label{eq:Hfull}
\hat{H}_{\text{tFH}}=&-J\sum_{i,\sigma} \left( \hat{c}_{i,\sigma}^{\dagger}\hat{c}_{i+1,\sigma} + \textrm{h.c}. \right) + U\sum_i \hat{n}_{i,\uparrow}\hat{n}_{i,\downarrow} \\
&+ \Delta \sum_{i,\sigma} i \hat{n}_{i,\sigma} \, .  \nonumber
\end{align}

\noindent In particular we will derive effective Hamiltonians corresponding to: (1) large tilt $\Delta \gg J,|U|$, (2) large interaction $|U|\gg J, \Delta$ and  (3) the resonant regime $|U|\simeq 2\Delta$.

\subsubsection{Large tilt limit: dipole conservation}

Here, we focus on the parameter regime $\Delta\gg \abs{U},J$ and derive an effective Hamiltonian using the high-frequency expansion (HFE) in the interaction picture.
The Hamiltonian in the interaction picture is time-periodic $ \hat H_I(t+\frac{2\pi}{\Delta})= \hat H_I(t)$.
According to Floquet theory~\cite{eckardt_high-frequency_2015,bukov_universal_2015} the unitary evolution generated by $\hat H_I(t)$ can be written as

\begin{equation}
\hat U_I(t,t_0)= e^{-i \hat K_{\textrm{eff}}(t)} e^{-i (t-t_0) \hat H_{\textrm{eff}}} e^{i \hat K_{\textrm{eff}}(t_0)},
\end{equation}

\noindent with a time-independent Floquet-gauge invariant Hamiltonian
$\hat H_{\textrm{eff}}$ and a gauge-dependent and time-periodic kick operator $ \hat K_{\textrm{eff}}(t)$.
It has been noticed that the first orders in the perturbative Schrieffer-Wolff (SW) transformation approach for static Hamiltonians (see e.g., \cite{bravyi_schrieffer-wolff_2011}), coincide with those in the HFE in the interaction picture (which provides the gauge-invariant effective Hamiltonian) \cite{mananga_introduction_2011,bukov_universal_2015}, with the SW generator given by the kick operators.
Following this approach, we obtain the effective Hamiltonian as a Floquet expansion in powers of $1/\Delta$ with $\hat H_{\textrm{eff}}=\sum_{n} \hat H_{\textrm{eff}}^{(n)}$ and $ \hat K_{\textrm{eff}}(t)=\sum_{n} \hat K_{\textrm{eff}}^{(n)}(t)$.
Up to third order the effective Hamiltonian is~\cite{goldman_periodically_2014,eckardt_high-frequency_2015,bukov_universal_2015,mikami_brillouin-wigner_2016}:

\begin{align} 
\label{eq:H3_dipole} 
\hat H_{\textrm{eff}}^{\mathrm{dip}}= &J^{(3)} \hat T_3
+ 2 J^{(3)} \hat T_{XY}  +U\Big( 1 -\frac{4 J^2}{\Delta^2}\Big)\sum_{i} \hat n_{i,\uparrow} \hat n_{i,\downarrow} \nonumber \\ 
&+ 2 J^{(3)} \sum_{i,\sigma} \hat n_{i,\sigma} \hat n_{i+1,\bar{\sigma}},
\end{align}

\noindent up to constant terms, where $\bar \sigma =\{\downarrow,\uparrow\}$ indicates the respective opposite spin of $\sigma=\{\uparrow,\downarrow\}$, $J^{(3)}=\frac{J^2U}{\Delta^2}$ and

\begin{align} \label{H3exp}
&\hat T_3= \sum_{i,\sigma}{ \hat c_{i,\sigma} \hat c_{i+1,\sigma} ^{\dagger} \hat c_{i+1,\bar{\sigma}} ^{\dagger} \hat c_{i+2,\bar{\sigma}}} +\text{h.c.},\\ 
&\hat T_{XY} =  \sum_{i,\sigma} \hat c_{i,\bar{\sigma}}^{\dagger} \hat c_{i+1,\bar{\sigma}} \hat c_{i+1,\sigma}^{\dagger}  \hat c_{i,{\sigma}}.
\end{align}

\noindent The kick-operator to third order is expressed as
\begin{align}
\hat{K}_{\textrm{eff}}(t)=&-i\frac{J}{\Delta}\sum_{i,\sigma}\big( \hat c^{\dagger}_{i,\sigma}e^{-it\Delta} \hat c_{i+1,\sigma} -\text{h.c.} \big) \nonumber\\
&-i \frac{JU}{\Delta^2} \sum_{i,\sigma}\big( \hat n_{i+1,\bar{\sigma}} - \hat n_{i,\bar{\sigma}} \big)\times \nonumber \\
&\ \quad \big(\hat c^{\dagger}_{i,\sigma}e^{-it\Delta} \hat c_{i+1,\sigma} - \text{h.c.}\big) 
\end{align}

\noindent and the time-evolution operator is approximated as
\begin{equation}
\hat U_I(t,t_0)\approx e^{-i \hat K_{\textrm{eff}}(t)}e^{-i (t-t_0) \hat H_{\textrm{eff}}} e^{i \hat K_{\textrm{eff}}(t_0)}.
\end{equation}

\noindent Rotating back to the Schrödinger picture, we find

\begin{align}
\hat{U}(t,t_0)=&e^{-i t \hat{H}_0}\hat{U}_I(t,t_0) e^{i t_0 \hat H_0 } \nonumber \\
&\approx e^{- \hat S} e^{-i (t-t_0) \big( \hat H_{\textrm{eff}}+ \hat H_0\big)}e^{ \hat S},
\label{eq:rotated_krylov}
\end{align}
\noindent where we have used the fact that $[ \hat H_{\textrm{eff}},\hat H_0]=0$ and that $ e^{-i t \hat H_0} \hat K_{\textrm{eff}}(t)e^{i t \hat H_0}=\hat K_{\textrm{eff}}(0)$ \cite{mananga_introduction_2011}, namely the product on the left hand side does not depend on time. 
Therefore, the Hamiltonian in the large-tilt limit can be approximated (up to higher-order terms) via
\begin{equation}
\hat H\approx e^{- \hat S}\big(\hat H_{\textrm{eff}}+ \hat H_0\big)e^{\hat S},
\end{equation}
\noindent taking the form of a perturbative SW transformation at third order in $J/\Delta$, with the SW generator given by $\hat S=i \hat K_{\mathrm{eff}}(0)$. 
We have thus obtained an effective Hamiltonian which conserves the dipole moment (or center of mass $\sum_{i, \sigma} i\hat{n}_{i, \sigma}$), with $\hat{T}_3$ in Eq.~\eqref{H3exp} the strongly-fragmented dipole-conserving Hamiltonian studied in \cite{sala_ergodicity_2020, khemani_localization_2020}, up to additional spin degrees of freedom. 
The fact that the hopping rate $J^{(3)}$ is proportional to the interaction strength highlights that interactions are necessary to generate dipole-conserving processes~\cite{taylor_experimental_2020} (pure off-diagonal non-interacting contributions destructively interfere at any order). $J^{(3)}$ agrees with the two particle picture~\cite{taylor_experimental_2020} yielding $J_{\mathrm{eff}}\propto \frac{U J^2}{\Delta^2 -U^2}$ with $|U| \ll \Delta$.
For CDW initial states of singlons, the connected dynamical sector $\mathcal{K}$ only represents a vanishing fraction of the whole (effective) symmetry sector $\mathcal{S}$, thus severely restricting the dynamics of the system.
The dipole-conserving processes in general involve the generation of doublons. This is, however, penalized by the Fermi-Hubbard on-site interaction in Eq.~\eqref{eq:H3_dipole} and therefore, we expect a slowing down of the dipole-conserving dynamics (see Sec.~\ref{sec:penalty}).  
The additional spin-exchange $\hat T_{XY}$ increases the connectivity, but cannot fully connect the whole dipole symmetry sector and the system remains fragmented.

\subsubsection{Large interaction limit}

We study the limit $\abs{U}\gg J, \Delta$ with $\left| \abs{U}-n\Delta\right| \neq 0$ for any $n\in \mathbb N$ to avoid possible resonances~\cite{de_roeck_very_2019}.
In this limit, the number of doublons $N_{\text{doub}}$ is effectively conserved up to times that scale exponentially in the interaction strength $U$ \cite{abanin_rigorous_2017,sensarma_lifetime_2010}.
Dealing with initial singlon configurations, we have $N_{\text{doub}}=0$ and assume a negligible fraction of dynamically-generated doublons after the quench. 
In this limit, the effective Hamiltonian provides non-trivial dynamics at first order in perturbation theory
\begin{align} 
\hat H_{\textrm{eff}}^U = &- J\sum_{i,\sigma}\big[(1- \hat n_{i,\bar{\sigma}}) \hat c_{i,\sigma}^{\dagger} \hat c_{i+1,\sigma}(1-\hat n_{i+1,\bar{\sigma}}) + \textrm{h.c.}\big] \nonumber \\
&+ \Delta \sum_{i,\sigma}  i \hat n_{i,\sigma}.
\label{eq:HU} 
\end{align}

\noindent Note that the dynamics generated by this Hamiltonian conserves the configuration of spins $\ket{\{\sigma_1,\dots,\sigma_N\}}$, with $\sigma_i=\{\uparrow, \downarrow\}$ and  the total particle number $N$.
The last term in Eq.~\eqref{eq:HU} equally couples to both spin degrees of freedom and the many-body states expressed in the particle-number basis factorize in terms of $N$ free Wannier-Stark localized spinless fermions with many-body wave function $\ket{\{i_1,\dots,i_N\}}$, with $i_i\in\{ -\frac{L}{2},\dots,0,\dots,\frac{L}{2}\}$ and fixed spin configuration $\ket{\{\sigma_1,\dots,\sigma_N\}}$~\cite{peres_finite-temperature_2000}. As a result the effective Hamiltonian [Eq.~\eqref{eq:HU}] takes the form
\begin{equation} 
\hat{H}_{\textrm{eff}}^U = - J\sum_{i}\big(\hat c_{i}^{\dagger} \hat c_{i+1} + \textrm{h.c.}\big)  +   \Delta\sum_{i} i \hat n_{i}.
\end{equation}

\noindent This has to be compared with the non-interacting Hamiltonian in Eq.~\eqref{eq:Hamiltonian} for $N=N_{\uparrow}+ N_{\downarrow}$ spinful fermions, which for a one-body observable like the imbalance gives exactly the same result. 
Higher-order terms at finite $U$ do not conserve the spin configuration $\ket{\{\sigma_1,\dots,\sigma_N\}}$. The leading terms in second-order perturbation are spin-exchange and longer-range hopping terms, as well as nearest-neighbors interactions $-2J^2/U\sum_{i,\sigma} \hat n_{i,\sigma} \hat n_{i+1,\bar \sigma}$, which lead to an interaction-induced decay of the imbalance to lower values compared to the non-interacting case at sufficiently long times ($t\sim U/J^2$). 

The experimental setup has a weak spin-dependent tilt ($\Delta_{\downarrow}- \Delta_{\uparrow}\approx 0.3J <J$), hence, the previous discussion provides a good approximation  for sufficiently strong $U$.
Only in the limit $\Delta_{\downarrow}- \Delta_{\uparrow} >J$, the effective Hamiltonian in Eq.~\eqref{eq:HU} does not map onto spinless fermions, because it depends on the spin configuration. This implies that the non-quadratic interaction terms, appearing in the hopping, have to be taken into account. 
This corresponds to two Stark ladders with different slopes constraining the mobility within each other.

\subsubsection{Resonant regime $|U| \simeq 2\Delta$}

The singlon CDW structure of the initial states makes the resonance $|U| \simeq 2\Delta$ more prominent in the dynamics than the one at $|U|=\Delta$, where any hopping process from the initial state would require an energy $\Delta$.
Consider the family of states for which $\hat H_0 = \Delta \sum_{i, \sigma} i \hat n_{i, \sigma} + 2\Delta \sum_i \hat n_{i,\uparrow} \hat n_{i,\downarrow}$ takes the same value. This defines a subspace, within which an effective Hamiltonian $ \hat H_{\textrm{eff}}$ with $[ \hat H_0, \hat H_{\textrm{eff}} ]=0$ can be obtained as an expansion in $\lambda=J/\Delta$. 
Such Hamiltonian can either independently conserve the dipole moment and the number of doublons or the sum of the two.
Using a Schrieffer-Wolf unitary transformation $e^{\lambda \hat S}$~\cite{bravyi_schrieffer-wolff_2011,lin_explicit_2017,abanin_rigorous_2017} with $\hat S=\sum_{n=0}\lambda^n \hat S_n$ up to an optimal order $n^*$, we can generate order-by-order an effective local Hamiltonian that is ``close'' to a block diagonal form with respect to $\hat H_0$
\begin{align}
&e^{\lambda \hat S_{n\leq n^*}} \hat  H e^{-\lambda \hat S_{n\leq n^*}}= \hat H^{(n^*)}_{\textrm{eff}} + \hat V_{{n\geq n^*}},
\end{align}
where $[\hat H_0, \hat V_{{n\geq n^*}}]\neq 0 $ with $\hat V_{{n\geq n^*}}$ exponentially small in $1/\lambda$~\cite{abanin_rigorous_2017, else_prethermal_2017, kuwahara_floquet-magnus_2016}.
In particular, we obtain the explicit form of the effective Hamiltonian to second order in $\lambda$: 
\begin{equation} 
\label{eq:H2_resonance}
\begin{split}
\hat H_{\textrm{eff}}^{\mathrm{res}}=& \hat H_0  + \frac{8J^2}{3\Delta} \sum_{i} \hat n_{i,\uparrow} \hat n_{i,\downarrow} - \frac{4J^2}{3\Delta} \hat T_{XY} + \frac{4J^2}{3\Delta}\hat H_D \\
&+ \frac{J^2}{\Delta} \hat T_1 - \frac{2J^2}{\Delta} \hat T_2 + \frac{2J^2}{3\Delta} \hat T^D_3 \, ,
\end{split}
\end{equation}

\noindent with
\begin{equation} \label{eq:contU2E}
\begin{split}
& \hat T_1= \sum_{i, \sigma}(1- \hat n_{i+2,\bar{\sigma}}) (1-2 \hat n_{i+1,\bar{\sigma}}) \hat n_{i,\bar{\sigma}} \hat c^{\dagger}_{i,\sigma} \hat c_{i+2,\sigma} + \textrm{h.c.},  \\ 
& \hat T_2=\sum_{i, \sigma}(1- \hat n_{i+2,\bar{\sigma}}) \hat n_{i,\sigma}c^{\dagger}_{i,\bar{\sigma}} \hat c_{i+1,\bar{\sigma}} \hat c^{\dagger}_{i+1,\sigma} \hat c_{i+2,\sigma}  + \textrm{h.c.} , \\
& \hat T^D_3= \sum_{i, \sigma}(\hat n_{i,{\sigma}}-\hat n_{i+2,\bar{\sigma}})^2(1-2(\hat n_{i+2,\bar{\sigma}} - \hat n_{i,{\sigma}})) \\
&\times \hat c_{i,\bar{\sigma}} \hat c^{\dagger}_{i+1,\bar{\sigma}}  \hat c^{\dagger} _{i+1,\sigma} \hat c_{i+2,\sigma}  + \textrm{h.c.}    \\
& \hat H_D= - 2 \sum_{i} \hat n_{i,\uparrow} \hat n_{i,\downarrow}  (\hat n_{i+1} - \hat n_{i-1}) -  \sum_{i,\sigma} \hat n_{i,\sigma} \hat n_{i+1,\bar{\sigma}}.
\end{split}
\end{equation}
\noindent The first term in the expansion of the SW generator $\hat S=\sum \lambda^n \hat S_n$ takes the form
\begin{equation}
\hat S_0=\sum_{i,\sigma} \big(1 -2 \hat n_{i,\bar\sigma} - \frac{2}{3} \hat n_{i+1,\bar\sigma} + \frac{8}{3} \hat n_{i,\bar \sigma} \hat n_{i+1,\bar\sigma} \big) \hat c_{i+1,\sigma}^{\dagger} \hat c_{i,\sigma}- \textrm{h.c.}. 
\end{equation}

\noindent Similar to the Hamiltonian $\hat H_{\textrm{eff}}^{\mathrm{dip}}$ [Eq.~\eqref{eq:H3_dipole}], $\hat H_{\textrm{eff}}^{\mathrm{res}}$ involves a ``dressed'' $\hat T_3^D$ term conserving both the dipole moment and the number of doublons independently, giving rise to doublon-assisted dipole conserving processes.
This is the diagonal part of $\hat T_3$ commuting with $\hat N_{\text{doub}}$.

\subsection{Fragmentation in the resonant regime $|U| \simeq 2\Delta$}
\label{sect:resfrag}

Here, we study the effective Hamiltonian $\hat H^{\mathrm{res}}_{\mathrm{eff}}$ in the resonant regime [Eq.~\eqref{eq:H2_resonance}]  regarding both its diagonal and off-diagonal terms in the number basis. While the diagonal terms cause a renormalized Fermi-Hubbard interaction and a shifted resonance, the off-diagonal terms result in a connectivity of the initial product states with other number basis states, causing strong fragmentation and  a finite steady-state imbalance.

\subsubsection{Renormalized Fermi-Hubbard interaction}

\begin{figure}[t!]
	\centering
	\includegraphics[width=3.3in]{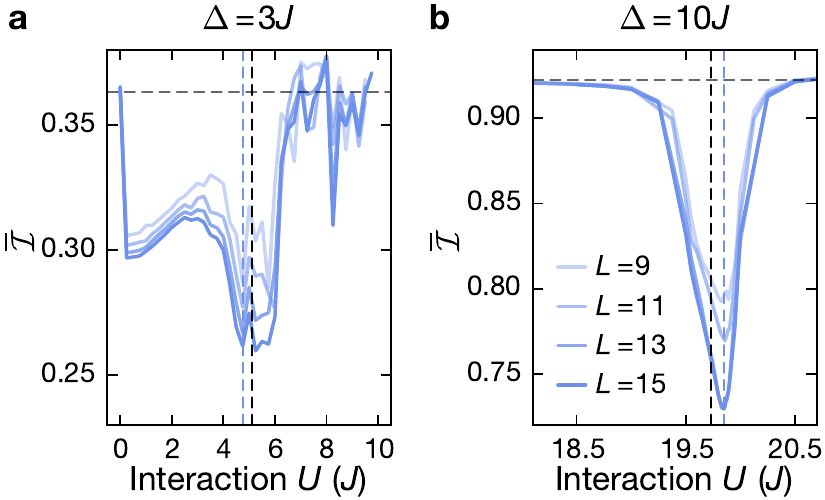}
	\caption{\textbf{Identification of the resonance $|U| \simeq 2\Delta$.} ED calculation of the time-averaged imbalance  $\overline{\mathcal{I}}=1/T \int_0^T \mathcal{I} dt$ for system sizes $L=9,11,13,15$ with increasing opacity and $T=1000\tau$. We use $\Delta_{\uparrow}=\Delta_{\downarrow}$. The horizontal dashed line shows the analytical value $\mathcal{J}_0(4J/\Delta)^2$ in the non-interacting case $(U=0)$ in the limit $T\to \infty$. The vertical black dashed line indicates the resonant point, including the second order correction $U_{\mathrm{res}}=2\Delta - 8J^2/(3\Delta)$.   \textbf{a} Time-averaged imbalance for $\Delta=3J$. Close to the minimum, we use a uniform grid with spacing $\delta U = 0.25 J$ and identify the lowest imbalance at $U=4.75J$ (blue dashed line). \textbf{b} Time-averaged imbalance for $\Delta=10J$. Close to the minimum we use a grid with steps $\delta U =0.01J$, allowing us to locate the minimum at $U=19.85 J$ (blue dashed line).}
	\label{fig:resonance}
\end{figure}

The diagonal terms of the effective Hamiltonian in Eq.~\eqref{eq:H2_resonance} add long-range interactions and renormalize the Fermi-Hubbard interaction such that the resonant point is shifted  for finite $\lambda$ according to $U + \frac{8J^2}{3\Delta} + \mathcal{O}(\frac{J^2}{\Delta})= 2\Delta$ and the overall resonance is broadened.
We numerically identify the resonance for large tilt $\Delta=10J$ (Fig.~\ref{fig:resonance}a) and intermediate tilt $\Delta=3J$ (Fig.~\ref{fig:resonance}b) using different system sizes $L=9,11,13,15$ probing the time-averaged imbalance $\bar{\mathcal{I}}(T)\equiv\frac{1}{T}\int_0^T dt\, \mathcal{I}(t)$. 
Fig.~\ref{fig:resonance} depicts a sharp resonance at strong tilt, while a rather broad feature is present at intermediate tilt. Away from $|U| \simeq 2\Delta$ in the large $U$ regime, we find that the system is Wannier-Stark localized. 
For both regimes, the numerical results are consistent with the analytic prediction for the shifted resonance to second order even after $1000 \tau$, as shown in Fig.~\ref{fig3}e in the main text.

\subsubsection{Strong Fragmentation}

\begin{figure}[t]
	\centering
	\includegraphics[width=3.3in]{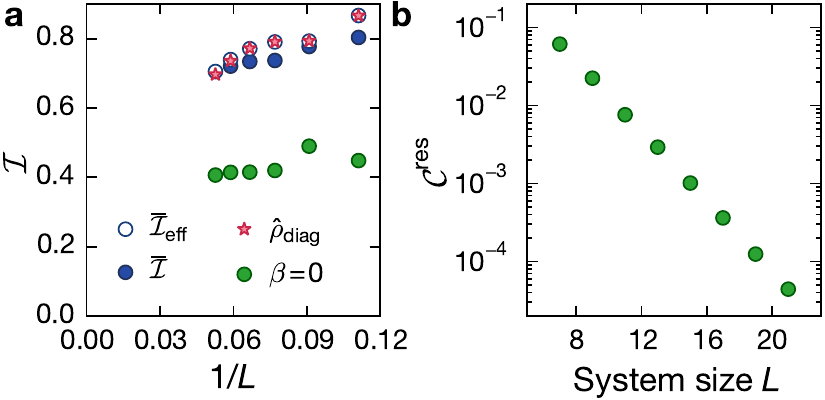}	
	\caption{\textbf{Imbalance in the large tilt limit $\Delta=10J$.} 
		\textbf{a} Finite size scaling of the long-time value of the imbalance calculated with the effective Hamiltonian in Eq.\eqref{eq:H2_resonance} using a time-averaged imbalance with $T=3000 \tau$ ($\overline{\mathcal{I}}_{\mathrm{eff}}$),  a diagonal ensemble ansatz ($\hat{\rho}_{\mathrm{diag}}$) and an infinite temperature prediction ($\beta = 0$). Additionally, the original Hamiltonian in Eq.~\eqref{eq:Hfull} is used to compare to the time-averaged imbalance calculated with $T=1000 \tau$ ($\overline{\mathcal{I}}$). All ED calculations were done with a Néel-ordered CDW initial state. 
		\textbf{b} System size scaling of the connectivity $\mathcal{C}^{\mathrm{res}}$ of the fragment $\mathcal{K}^{\mathrm{res}}$, capturing the Néel-ordered CDW initial state within the full Hilbert space $\mathcal{H}$, restricted to quarter filling and zero magnetization. }
	\label{fig:Krylov}
\end{figure}

The off-diagonal terms of the effective Hamiltonian in Eq.~\eqref{eq:H2_resonance} consist of three different kinds of correlated hoppings $\hat{T}_1, \hat{T}_2, \hat{T}^D_3$ (see Fig.~\ref{fig4}d in the main text) and all hopping rates scale as $J^2/\Delta$. 
Since $[\hat H^{\mathrm{res}}_{\textrm{eff}},\hat H_0]=0$, $\hat H_0$ becomes a new global quantum number fixed by the initial configuration, i.e. the linear combination of the dipole moment and the number of doublons is perturbatively conserved. 
Generically, after fixing this new global quantum number, the corresponding symmetry sector $\mathcal{S}$ is fully connected by the action of the effective Hamiltonian and the Krylov subspace, hosting the initial state, agrees with the global symmetry sector $\mathcal{S}$.
In contrast, we realize that this is not the case for the effective Hamiltonian in Eq.~\eqref{eq:H2_resonance}. Here, the symmetry sector decomposes into exponentially many disconnected fragments $\mathcal{K}^{\mathrm{res}}$ and the initial state remains trapped within such a fragment without exploring the whole symmetry sector.

For the subsequent analysis, we use a Néel-ordered CDW initial state, expected to show the strongest interaction effects and fastest dynamics.
The correlated hoppings $\hat T_1, \hat T_2, \hat T_3^D$ of the effective Hamiltonian connect the initial state with a set of states defining the fragment $\mathcal{K}^{\mathrm{res}}$. 
Similarly to the finite-time connectivity $\mathcal{C}_\epsilon$ of the numerical fragment (Fig.~\ref{fig1}, Fig.~\ref{fig4} in the main text), we define the connectivity $\mathcal{C}^{\mathrm{res}}=\mathrm{dim}(\mathcal{K}^{\mathrm{res}})/\mathrm{dim}(\mathcal{H})$ as the ratio between the dimension of the fragment $\mathrm{dim}(\mathcal{K}^{\mathrm{res}})$ and the Hilbert space $\mathrm{dim}(\mathcal{H})$, which is restricted to quarter filling and zero magnetization.
Experimentally, we do not realize Néel-ordered CDW states, but the connectivity of our initial state with random CDW spin-sector is the same as for the Néel-ordered CDW state. 
In Fig.\ref{fig:Krylov}b, we show the system size scaling of the connectivity and find that it vanishes exponentially in the thermodynamics limit as expected in the regime of strong fragmentation (The same scaling holds for the connectivity of the fragment within the symmetry sector $\mathcal{S}$.)~\cite{sala_ergodicity_2020,moudgalya_thermalization_2019,khemani_localization_2020, rakovszky_statistical_2020}.

In Fig.~\ref{fig:Krylov}a we analyze the system size scaling of both the infinite temperature (within the fragment containing the initial state) and diagonal ensemble predictions for the imbalance, obtaining a positive result in both cases for system sizes $L=9,11,13,15,17,19$ with no clear convergence towards zero imbalance in the thermodynamic limit. 
The scaling of the infinite temperature prediction suggests a finite value even in this limit. 
This apparent finite imbalance for $\hat H_{\textrm{eff}}^{\mathrm{res}}$ could be interpreted as follows: Given an initial state that breaks even-odd sublattice symmetry, most dynamical processes in Eq.\eqref{eq:H2_resonance}, except those generated by $\hat T^D_3$, do only transport particles in one of the sublattices. 
Thus, most states within the fragment have positive imbalance in agreement with the positive infinite temperature value. 
This explanation is in line with the observed ergodicity-breaking in dipole-conserving systems, where a finite value of the autocorrelation was observed even at infinite temperatures~\cite{sala_ergodicity_2020}.

In Fig.~\ref{fig:Krylov}a, simulations with the exact Hamiltonian $\hat H_{\text{tFH}}$ [Eq.~\eqref{eq:Hfull}] for $\Delta=10J$, $U=19.85 J$ agree well with the results of the effective Hamiltonian [Eq.~\eqref{eq:H2_resonance}] even up to remarkably long times $T\sim 10^3 \tau$. 
Consistent with a perturbative expansion in $\lambda$, which neglects higher order terms in the effective Hamiltonian,  it yields a systematically larger imbalance compared to the original Hamiltonian.
Since the conservation law, i.e. the linear combination of the dipole moment and the number of doublons, only holds perturbatively, one would expect that it is valid only up to a certain timescale.

\subsubsection{Numerical fragment $\mathcal{N}_\epsilon$ versus fragment $\mathcal{K}^{\mathrm{res}}$}

\begin{figure}[t]
	\centering
	\includegraphics[width=3.3in]{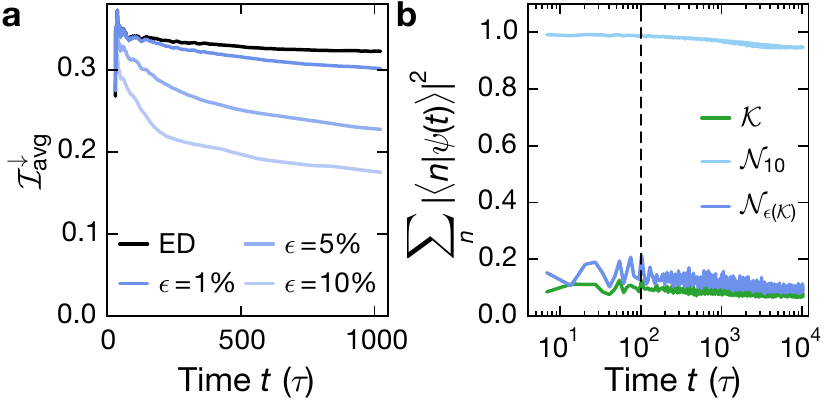}	
	\caption{\textbf{Numerical fragment $\mathcal{N}_\epsilon$.} Both figures use $U=5J$ and $\Delta_\downarrow=\Delta_\uparrow=3J$. 
		\textbf{a} Imbalance time traces calculated with different sets of states  $\mathcal{N}_\epsilon$ and ED for $U=5J$ and $\Delta=3J$. $I^\downarrow_{\mathrm{avg}}$ is calculated using a cumulative sum to reduce fluctuations; $L=11$. 
		\textbf{b} Contribution of the set of states $\ket{n}$ in the numerical fragments $\mathcal{N}_{\epsilon ( \mathcal{K}^{\mathrm{res}})}$ and $\mathcal{N}_{10}$ and the fragment $\mathcal{K}^{\mathrm{res}}$ to the time-evolved initial state $\ket{\psi(t)}$. $T_\mathcal{N}=100 \tau$ (dashed line).  }
	\label{fig:numerical_fragment}
\end{figure}

In Fig.~\ref{fig:numerical_fragment}a, we investigate how well the imbalance can be captured, when using only the states within the numerical fragment. These states correspond to a small fraction compared to the states within the full Hilbert space and this fraction was found to vanish in the thermodynamic limit (Fig~\ref{fig4}b). We show imbalance time traces, calculated with a cumulative sum to reduce fluctuations and compare traces with different cut-off values $\epsilon$ to the exact numerical result, which we obtained with ED. We find that for $U=5J$ and $\Delta=3J$, already with a cut-off $\epsilon=1 \%$ we can reproduce the exact result well, larger cut-off values result in a deviation, which becomes more pronounced at later times.
In Fig.~\ref{fig:numerical_fragment}b the overlap of the states $\ket{n}$ in different numerical fragments and in the Krylov subspace $\mathcal{K}^{\mathrm{res}}$ with the time evolved initial state $\ket{\psi (t)}$ is analyzed by calculating $\sum_n |\braket{n | \psi (t)}|^2$. While the overlap in our parameter regime is poorly captured by the states in the Krylov subspace $\mathcal{K}^{\mathrm{res}}$ (as expected because these states best describe the time evolution only in the limit $|U|=2\Delta \gg J$), we can find the proper states by choosing a small enough $\epsilon$. For $\mathcal{N}_1$ we get $\sum_n |\braket{n | \psi (t)}|^2 \approx 1$, which shows a very weak decay even up to $t=10^4 \tau$. Note that it is crucial to choose enough states for the numerical fragment. If we choose the same number of states as used in the Krylov subspace $\mathcal{K}^{\mathrm{res}}$ for the numerical fragment $\mathcal{N}_{\epsilon ( \mathcal{K}^{\mathrm{res}})}$, we cannot capture the time evolved initial state $\ket{\psi(t)}$ well.

\subsection{Constrained dynamics in the presence of higher order terms and on-site interactions}

\label{sec:penalty}

Dynamics caused by fragmentation is captured by effective Hamiltonians and is therefore a transient phenomenon. The perturbative  derivation of the effective Hamiltonian neglects higher-order terms which are known to eventually couple different fragments and symmetry sectors, such that the dynamics no longer solely occur within a certain fragment. 
Estimating the time scales, which capture the dynamics caused due to fragmentation, requires a detailed analysis of both the diagonal and off-diagonal terms of the effective Hamiltonian. Note that the off-diagonal term $\hat T_3$, occurring at a rate $J^{(3)}=\frac{J^2 U}{\Delta^2}$ in the dipole conserving limit ($\Delta/J \rightarrow \infty$, Eq.~\eqref{eq:H3_dipole}), requires the production of doublons. Creating a doublon is, however, penalized by the diagonal Fermi-Hubbard interaction with strength $\sim U$. Therefore, an initial state consisting of a CDW of singlons without doublons remains frozen for exponentially long times $t\geq e^{c(\Delta/ J)^2}$, analogously to the stability of doublons in the repulsive Fermi-Hubbard model in the $U\gg J$ regime~\cite{sensarma_lifetime_2010, abanin_exponentially_2015}. This effectively gives rise to a fragmentation not only due to the conservation law of the respective effective Hamiltonian, but additionally due to the conservation of the doublon number~\cite{rakovszky_statistical_2020}. 
A similar argument can be made for the time scale on which higher-order off-diagonal terms, coupling different fragments, become effective and eventually destroy fragmentation. We will give a brief outline here for the dipole conserving limit, where higher order terms are easier to capture. 
These terms add longer-range processes to the effective Hamiltonian and in general order-$n$ terms  generate longer range-$n$ processes whose effective hopping rate scales as $J^{(n)}\sim J^{2k}U^{n-2k}/\Delta^{n-1}$ for some $k$. Any even order vanishes due to destructive interference: For every process started by a particle hopping to the left, there exists another process with a particle hopping to the right, thus contributing with opposite signs. 
The hopping rate of the next non-vanishing fifth-order scales as  $J^{(5)}\sim J^4U/\Delta^4$.

\begin{figure}[t]
	\centering
	\includegraphics[width=3.3in]{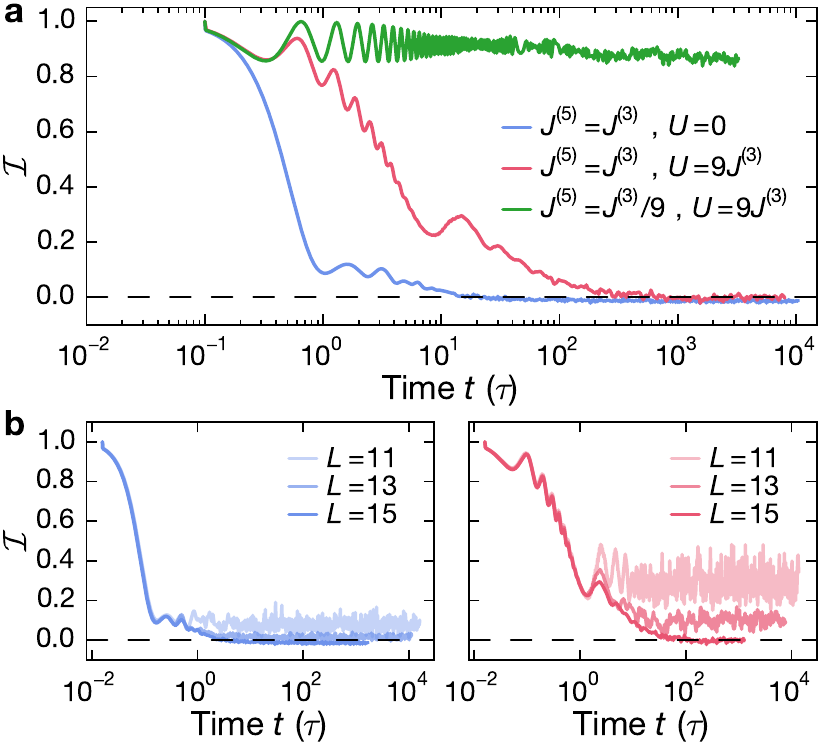}
	\caption{\textbf{Role of higher-order diagonal and off-diagonal terms.} ED calculation of the imbalance for the Hamiltonian $\hat H= J^{(3)} \big( \hat T_3 + \hat T_{XY} \big) + J^{(5)}\big(\hat T_4 + \hat T_5\big) +U \hat N_{\text{doub}}$. \textbf{a} Imbalance for $J^{(5)}=J^{(3)}$, $U=0$ (blue), $J^{(5)}=J^{(3)}$, $U=9J^{(3)}$ (red), and $J^{(5)}=J^{(3)}/9$, $U=9J^{(3)}$ (green) for system size $L=15$. \textbf{b} Finite size scaling of the imbalance for $J^{(5)}=J^{(3)}$, $U=0$ (left) and for $J^{(5)}=J^{(3)}$, $U=9J^{(3)}$ (right). In both cases, we use $L=11,13,15$ and increasing opacity corresponds to increasing system size. }
	\label{fig:artificial}
\end{figure}

\begin{figure*}[t!]
	\centering
	\includegraphics[width=\textwidth]{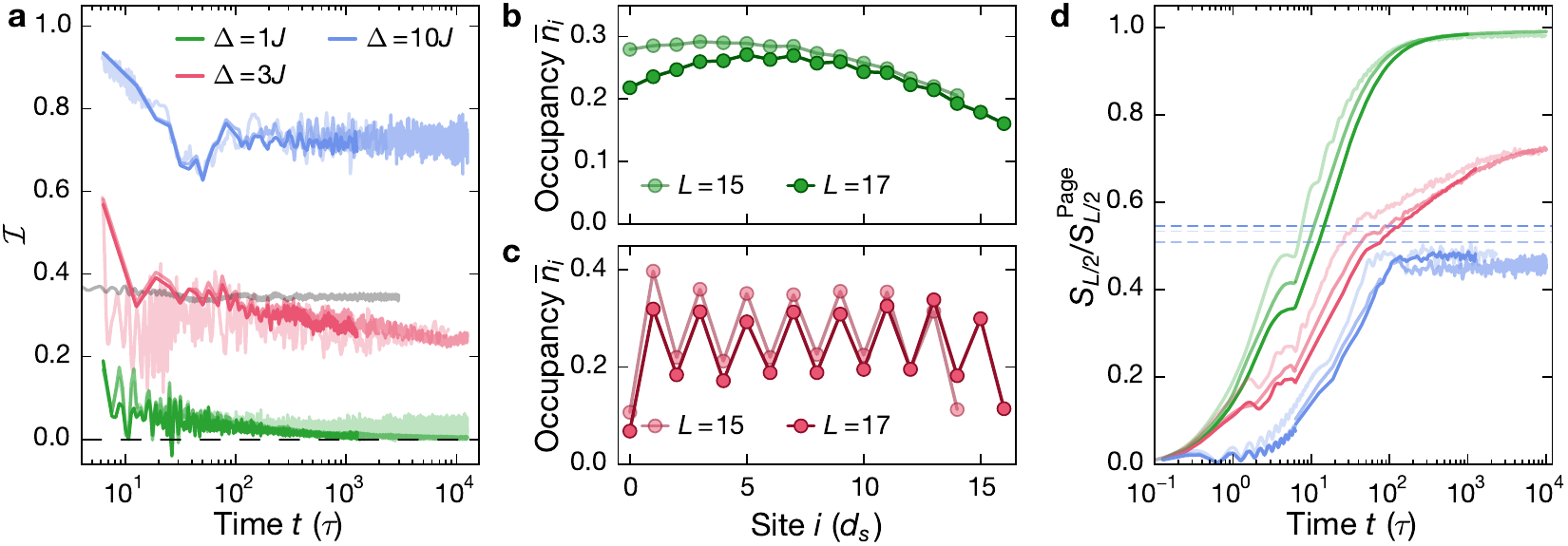}	
	\caption{\textbf{Finite-size scaling analysis of imbalance, entanglement entropy and occupancy.} 
		\textbf{a} Long-time behavior of imbalance $\mathcal{I}$ for system sizes $L=13,15,17$ and $(\Delta=10J, U=19.85J)$ (blue), $(\Delta=3J, U=4.75J)$ (red) and system sizes $L=12,14,16$ for $(\Delta=1J, U=4.75J)$ (green). The grey line corresponds to a simulation of the imbalance according to the effective Hamiltonian $\hat H_{\mathrm{eff}}^{\mathrm{res}}$ [Eq.\eqref{eq:H2_resonance}, Eq.~\eqref{eq:rotated_krylov}] for $L=15$ and $\Delta=3J$ up to $3000 \tau$. Fluctuations in the data are reduced by using a running average with a time-window of $10 \tau$.
		\textbf{b} ,\textbf{c} Time-averaged on-site occupancy $\overline{n}_i=10/T \int_{0.9T}^T n_i dt$ for system sizes $L=15, 17$ and \textbf{b} $(\Delta=1J, U=4.75J)$ and  \textbf{c}  $(\Delta=3J, U=4.75J)$ . The time average was performed with $T=12600\tau $ for $L=15$ and $T=1260 \tau$ for $L=17$.   
		\textbf{d} Long-time behavior of the half-chain entanglement entropy $S_{L/2}$ normalized to the Page value $S_{L/2}^{\mathrm{Page}}$ within the $(N_{\uparrow}, N_{\downarrow})$ symmetry sector for the same parameters as in (a) and system sizes $L=13,15,17$. The dashed horizontal lines shows the entanglement entropy of a random state within the fragment $\mathcal{K}^{\mathrm{res}}$ containing the Néel-ordered CDW initial state. Increasing opacity corresponds to increasing system size. All calculations were done using ED.}
	\label{fig:scaling}
\end{figure*}

Using a qualitative Kato-Bloch perturbative approach \cite{soliverez_an_1969, kato_on_1949}, which is easier to handle than a Schrieffer-Wolf transformation or a Floquet expansion~\cite{prato_a_1997} for higher-order terms, two terms emerge at fifth order in the dipole conserving limit: a $5$-local Hamiltonian $\hat T_5=\sum_{i,\sigma} \left( \hat c_{i,\sigma} \hat c^{\dagger}_{i+2,\sigma} \hat c^{\dagger}_{i+2,\bar \sigma} \hat c_{i+4,\bar \sigma} + \textrm{h.c.} \right) $, 
with two opposite spins hopping to an intermediate site, requiring the creation of a doublon in the central site.
A $4$-local term $\hat T_4$ similar to the $\hat H_4$ Hamiltonian studied in \cite{sala_ergodicity_2020}
$\hat T_4=\sum_{i,\sigma} \left(\hat c_{i,\sigma} \hat c^{\dagger}_{i+1,\sigma} \hat c^{\dagger}_{i+2,\bar \sigma} \hat c_{i+3,\bar \sigma} + \textrm{h.c.} \right) $, which populates nearby sites with opposite spin, thus interacting via the nearest-neighbor interaction appearing at third order. 
We now consider the time-evolution of an Néel-ordered CDW initial state for system size $L=15$ with the toy model Hamiltonian $\hat H= J^{(3)} \big( \hat T_3 + \hat T_{XY} \big) + J^{(5)}\big(\hat T_4 + \hat T_5\big) +U \hat N_{\text{doub}}$ using $J^{(3)}=1$ as unit of energy. Here, $\hat N_{\text{doub}}$ measures the number of doublons. In Fig.~\ref{fig:artificial}a we clearly observe an exponential decay of the imbalance for $J^{(3)}, J^{(5)}\sim O(1)$ and $U=0$ in agreement with the results in Refs.~\cite{sala_ergodicity_2020,khemani_localization_2020}.
The decay time scale increases strikingly, when adding on-site interactions such that $J^{(3)}=J^{(5)}=1$, $U=9$ corresponding to a ratio $U/J^{(3)}=9$ in the perturbative expansion, which is consistent with $\Delta=3J$ although the higher-order term is still unrealistically large ($J^{(5)}=J^{(3)}$).

A more realistic regime is captured with $J^{(5)}=J^{(3)}/9$ and $U=9$ due to the perturbative scalings. Here, the imbalance clearly stays finite on our time scales.
Thus, the energy penalty given by the on-site interaction has a drastic effect on the decay of the imbalance  caused by higher order terms, slowing down the dynamics tremendously. 
Fig.~\ref{fig:artificial}b and Fig.~\ref{fig:artificial}c show a finite-size scaling in the regimes $J^{(5)}=J^{(3)}$ with $U=0$ and $J^{(3)}=J^{(5)}$ with $U=9$, clearly indicating that large system sizes are necessary to capture the correct steady-state imbalance.

Unlike the previous regime, at perfect resonance $U_{\mathrm{res}} = 2\Delta \gg J$, neither the lowest-order dynamical processes generated by $\hat H^{\mathrm{res}}_{\mathrm{eff}}$ in Eq.~\eqref{eq:H2_resonance} nor in general higher-order terms, are energetically suppressed. Thus, at a time scale given by the fourth-order term $t \propto \Delta^3/J^4$, fragmentation phenomena are expected to breakdown with the result that imbalance decays. Note that the third-order and in general any odd-order term vanishes due to the CDW initial state, requiring an even number of hoppings for a resonant exchange between tilt  and interaction energy. Locating such a resonant point (at finite $\Delta$) requires fine-tuning: every order in perturbation theory gives a diagonal contribution renormalizing the Fermi-Hubbard interaction. As numerically shown in Fig.~\ref{fig:resonance}a, this is even more subtle at lower values of the tilt. In general, we expect a finite detuning from the resonance, which can be comparable to higher-order contributions, thus 'shielding' the fragmentation of the lowest order Hamiltonian and slowing down the dynamics.

\subsection{Scaling analysis of the steady-state imbalance and the entanglement entropy}

Here, we study the system-size scaling of the long-time dynamics in a clean system without spin-dependent tilt and harmonic confinement for a large range of tilts: We choose a weak tilt $\Delta=1J$, an intermediate tilt $\Delta=3J$ and a large tilt $\Delta=10J$.
We focus on the dynamics close to the resonant point $|U|\approx 2\Delta$, and consider an initial Néel-ordered CDW state. 
This state has a symmetric charge distribution with respect to the center site and thus its dipole moment coincides with that of a homogeneous charge distribution. 
In Fig.~\ref{fig:scaling}a we show numerical simulations of the imbalance $\mathcal{I}$ up to late times for different system sizes. In the large tilt regime, we find a stable imbalance for all system sizes, whereas the intermediate and weak tilt regime show an imbalance decay. This decay is very weak in the intermediate tilt regime and a conclusive answer on whether and at what timescale the imbalance decays to zero cannot be given. In contrast, the imbalance calculated with the effective Hamiltonian $\hat H_{\mathrm{eff}}^{\mathrm{res}}$ [Eq.\eqref{eq:H2_resonance}, Eq.~\eqref{eq:rotated_krylov}] is stable (grey shaded trace in Fig.~\ref{fig:scaling}a), as expected due to the absence of higher-order terms in the perturbative construction. Additionally, we find that the imbalance weakly scales down with system size. For small tilts, we clearly observe a decay of the imbalance to zero for large enough system sizes. 

Note that, while we used  $L=13,15,17$ for the intermediate and large tilt regime to minimize edge effects with an unoccupied odd site at the left and the right end of the system, we choose $L=12,14,16$ for the weak tilt regime. In this regime, the initial CDW relaxes to a potentially thermal density distribution and such a distribution only has zero imbalance for an equal number of even and odd sites. Additionally, the breathing amplitude of the dynamics for $\Delta=1J$ is four sites and boundary effects cannot be easily prevented by including an empty site at the edges. 
We confirm in Fig.~\ref{fig:scaling}b that the on-site occupancy shows no more memory of the initial CDW order in the regime of weak tilt, consistent with a zero imbalance. For the intermediate tilt in Fig.~\ref{fig:scaling}c, we clearly find a remaining CDW order.

In Fig.~\ref{fig:scaling}d we show numerical simulations of the half-chain entanglement entropy $S_{L/2}$, normalized to the Page value $S_{L/2}^{\textrm{Page}}$~\cite{page_average_1993, vidmar_entanglement_2017, de_tomasi_multifractality_2020}. The Page value $S_{L/2}^{\textrm{Page}}$ is the half-chain entanglement entropy of a pure random state within the symmetry sector fixed by particle number $N_{\uparrow},N_{\downarrow}$. The half-chain entanglement entropy of an ergodic system at infinite temperature is in general expected to reach $S_{L/2} = S_{L/2}^{\textrm{Page}}$. 
In the weak tilt regime, the half-chain entanglement entropy converges towards the thermal Page value for large enough system sizes, which is consistent with a lack of memory of the initial state as observed with the imbalance and the on-site occupancy.
For an intermediate tilt, we observe a sub-thermal entanglement entropy, growing only very slowly at late times, which is consistent with the finite imbalance up to the latest times accessible in the simulations. 
For large tilt, the entanglement entropy reaches a plateau, which slightly depends on the system size. This saturation value of the entanglement entropy is slightly smaller then the entanglement entropy of a random state within the fragment $\mathcal{K}^{\mathrm{res}}$ (blue dashed lines for the different system sizes) in which the initial state is contained.

\begin{figure}[t!]
	\includegraphics[width=3.3in]{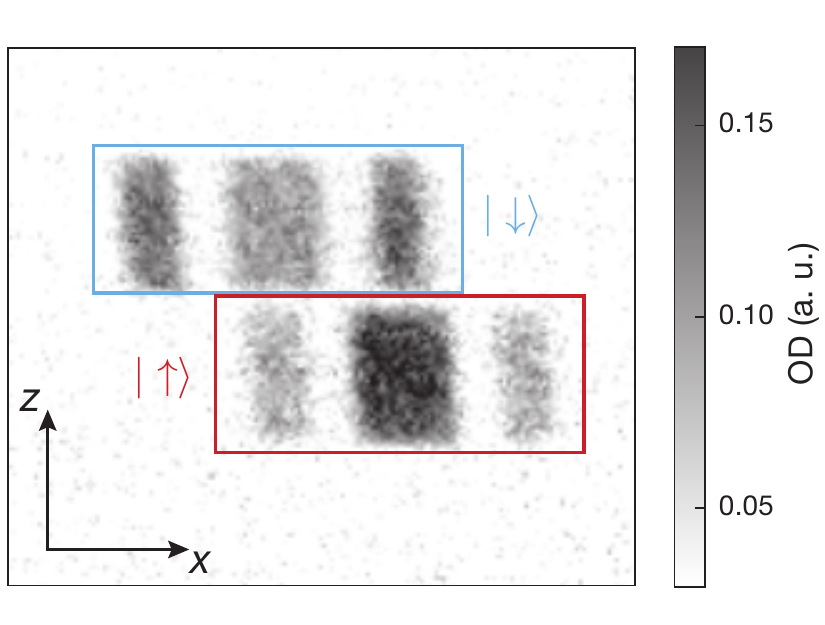}
	\caption{\textbf{Exemplary raw image.} The two spin states $\ket{\uparrow}$, $\ket{\downarrow}$ are spatially separated using Stern-Gerlach resolved time-of-flight imaging to extract the imbalances of each component $\mathcal{I}^\sigma$ independently. The populations of first and third band are extracted via a pixel sum of the optical density (OD).}
	\label{fig:bandmapping}
\end{figure}

\subsection{Experimental sequence}
\label{sec:sequence}

\subsubsection{General description}

\begin{figure*}[t!]
	\includegraphics[width=\textwidth]{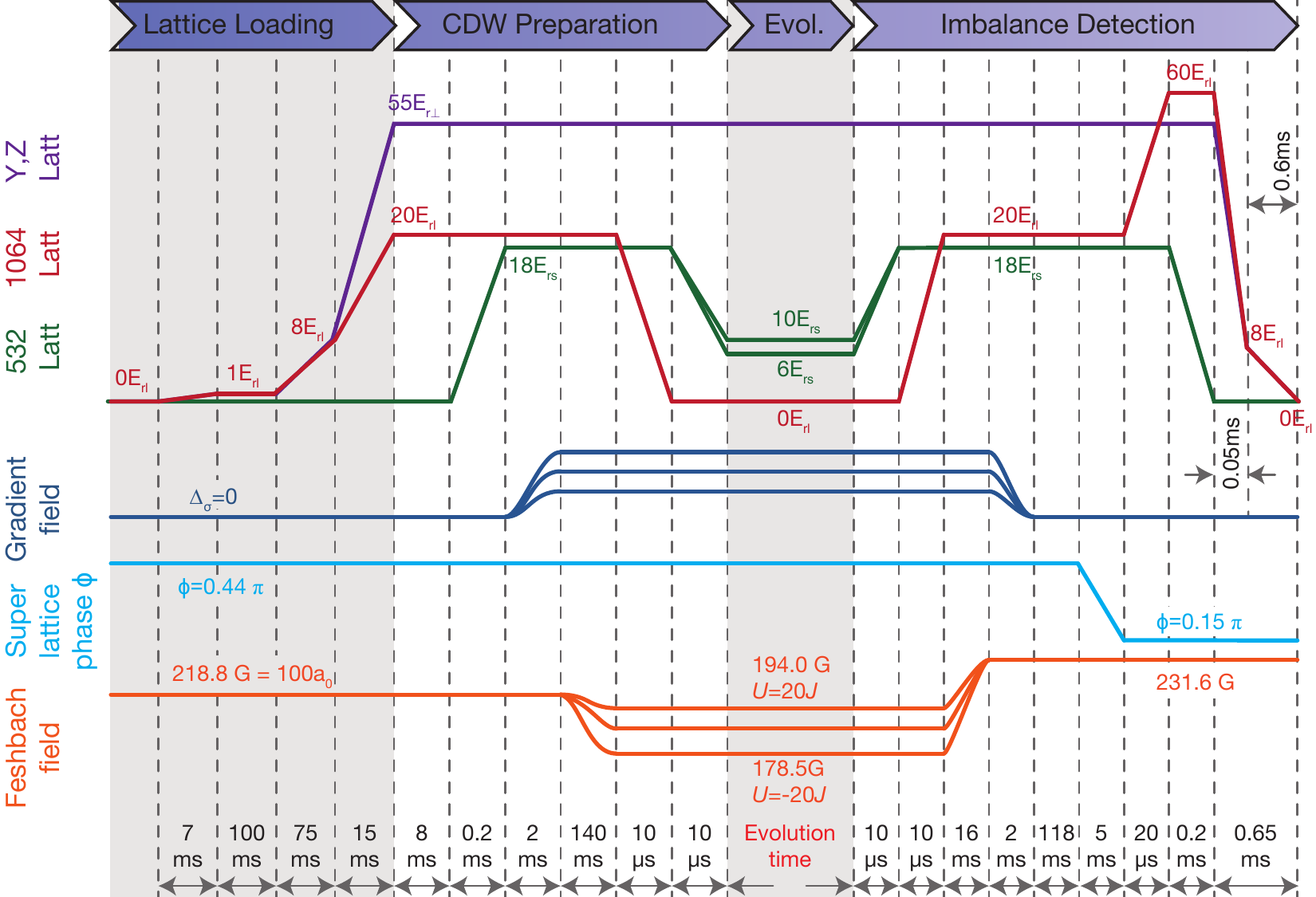}
	\caption{\textbf{Timing protocol of the experimental sequence.} Schematic showing the lattice depths, the superlattice phase, Feshbach field ramps and the gradient field ramps for loading, CDW preparation, time evolution and detection of the imbalance.}
	\label{fig:sequence}
\end{figure*}

We create a degenerate Fermi gas of $^{40}\mathrm{K}$ atoms in a crossed beam dipole trap. While the non-interacting traces are measured with a spin-polarized gas with all atoms in the state $\ket{\downarrow} = \ket{F=9/2, m_F = -9/2}$, we work with an equal mixture of the states $\ket {\uparrow} = \ket{F=9/2, m_F = -7/2}$ and $\ket{\downarrow}$ when studying interacting dynamics.  Details of both the cooling sequence and the preparation of the spin-polarized gas can be found in a previous publication~\cite{schreiber_observation_2015}. The initial state preparation starts with repulsively loading the atoms at a scattering length of $a=100 a_{0}$ into a three-dimensional (3D) optical lattice by a series of linear ramps (see Fig.~\ref{fig:sequence}). The scattering length can be tuned with a Feshbach resonance between the two states of the spin mixture centered at $202.1 \,\mathrm{G}$. While the repulsive scattering length suppresses the formation of doubly-occupied sites (doublons) during the loading, we extinguish any residual doublons by applying a $\SI{100}{\micro \second}$ off-resonant light pulse right after loading the deep lattice~\cite{scherg_nonequilibrium_2018}. The off-resonant light pulse results in light assisted collisions, which remove doublons without harming atoms on singly-occupied sites (singlons). Afterwards, we end up with singlons in an array of 1D tubes to which the dynamics is restricted on the observed timescales due to the deep orthogonal lattices. Using Gaussian fits to the atom cloud in the lattice, we characterize the $4 \sigma$ width of the central tubes to $L_\text{exp}=290$ sites. Along the $y$ direction and the $z$ direction, we populate about $150$ sites and $22$ sites, respectively.

We create the tilted lattice by applying a magnetic field gradient with a single coil. This coil, however, not only creates a field gradient, but additionally a strong homogeneous magnetic field component up to $110 \,\mathrm{G}$ is present, which adds to the homogeneous Feshbach field created by a pair of Helmholtz coils. Independent control of both the tilt and the interaction strength requires a tilt-dependent reduction of the Feshbach field and leads to extended wait times to reach stable currents through the coils. We use a wait time of $\SI{140}{\milli \second}$ before time evolution and another $\SI{136}{\milli \second}$ after time evolution before the band transfer to ensure a stable Feshbach field (see Fig.~\ref{fig:sequence}). Residual dynamics within a 1D tube during the wait time are suppressed by holding the atoms in strongly tilted double wells. All experiments throughout this work employ Stern-Gerlach resolved absorption imaging after $\SI{6.4}{\milli \second}$ time-of-flight (Fig.~\ref{fig:bandmapping}).

\subsubsection{CDW preparation and spin-resolved imbalance readout}

After creating an array of 1D tubes in a deep 3D optical lattice, we ramp up the short lattice $\lambda_s =  \SI{532}{\nano \meter}$ in addition to the long lattice $\lambda_l = \SI{1064}{\nano \meter}$ along the $x$ direction at a superlattice phase of $\phi=0.44\pi$ within $\SI{200}{\micro\second}$. Here, we use the convention that a symmetric double-well potential is realized for $\phi=k\cdot\pi$, with $k=\mathbb{Z}$. This creates strongly tilted double wells with one atom located on the low-energy site of each double well (even site), while the high energy site (odd site) is empty. This charge-density wave (CDW) state is time evolved in all experiments throughout this work. After time evolution, we apply a band transfer technique in the superlattice~\cite{sebby-strabley_lattice_2006, foelling_direct_2007}, which maps atoms on odd sites (high-energy site of each double well) into the third band of the long lattice, while atoms on even sites remain in the first band. Here, we require a different superlattice phase of  $\phi=0.15\pi$. Afterwards we perform bandmapping and Stern-Gerlach resolved absorption imaging to evaluate the spin-resolved imbalance. The Stern-Gerlach gradient and the magnetic field gradient during time evolution are created by the same coil.

A large enough spatial separation of the two spin states $\ket{\uparrow}$ and $\ket{\downarrow}$ during Stern-Gerlach resolved bandmapping is achieved by applying a Landau-Zener sweep before the band transfer to convert atoms from $\ket{\uparrow} = \ket{F=9/2, m_F=-7/2}$ to $\ket{\rightarrow}=\ket{F=9/2,m_F=-5/2}$. This sweep is performed at a set magnetic field of $231.6\,\mathrm{G}$, corresponding to the zero crossing of the Feshbach resonance between the two states  $\ket{\downarrow}$ and $\ket{\rightarrow}$, centered around $224.2\,\mathrm{G}$. We perform a linear frequency ramp with a duration of $\SI{10}{\milli \second}$ centered at $\SI{51.87}{\mega \hertz}$ with a deviation of $\SI{1}{\mega\hertz}$. Switching off interactions between these two states ensures the absence of interband oscillations after the transfer to the third band. Additionally, non-interacting bandmapping results in sharper edges of the absorption images and improves the accuracy of the imbalance measurement. A sample raw image used for data acquisition is shown in Fig.~\ref{fig:bandmapping}.

Measuring a perfect imbalance equal to one can be compromised by many artifacts such as an imperfect initial state preparation and a finite transfer efficiency of the population on odd sites into the third band. In order to calibrate these imperfections, we take two different sets of images. The first set measures the highest possible initial imbalance (around 0.92(2)) with no evolution time. The second set measures the imbalance after $\SI{25}{\milli \second}$ evolution time without tilt, which is supposed to yield zero. We then calculate a matrix that maps the measured imbalances for these two sets to 1 (first set) and 0 (second set). In particular, we have to determine a $2\times 2$-matrix $A^\sigma$, for each state $\sigma=\uparrow, \downarrow$, which satisfies

\begin{equation}
\begin{pmatrix}
n_{e,1}^\sigma & n_{e,2}^\sigma \\
n_{o,1}^\sigma & n_{o,2}^\sigma
\end{pmatrix}
A^\sigma = 
\begin{pmatrix}
1 & 0.5 \\
0 & 0.5
\end{pmatrix}.
\end{equation}

\noindent Here, $n_{e,i}^\sigma$ ($n_{o,i}^\sigma$) denote the relative atom number on even (odd) sites for the respective spin state and $i=1,2$ refers to the imbalance in the respective set (first or second set). This matrix is then used to rescale the measured imbalance for each spin component.

\subsection{Creating a homogeneous potential}

\label{sec:hom_pot}

Before loading into the lattice the atoms are confined by three dipole trap beams. The horizontal beams along $x$ (dynamics) and $y$ are elliptical ($30\times \SI{300}{\micro\meter}$) and the vertical beam along the $z$-direction is circular with a Gaussian beam waist of $\SI{150}{\micro\meter}$. All optical lattice beams have the same size as the $z$-dipole trap and are blue-detuned, thus providing an anti-trapping potential. A flat potential during the Bloch oscillations can be achieved by compensating the confinement of the vertical dipole trap with the anti-confinement of the optical lattices. The horizontal traps should only marginally contribute to the total confinement and we find the optimal configuration if the $x$-dipole trap is switched off and the $y$-dipole trap provides a very weak confinement during the time evolution. The confinement is optimized by fixing the time to four Bloch cycles ($t=4T_\downarrow$) and scanning the dipole trap strength on maximal imbalance. This method works because the confinement does not lead to a frequency change of the oscillations.

\subsection{Creating a linear potential}

The tilt is created by applying a magnetic field gradient. The energy $E^\sigma$ of each state $\ket{\sigma}$ in the $F=9/2$ hyperfine ground-state manifold in the presence of a magnetic field can be analytically calculated using the Breit-Rabi formula and a magnetic field gradient  results in a linear potential $\Delta_{\sigma}i$  according to $\Delta_{\sigma}=\frac{dE^\sigma}{dB}\partial_x B$. The first factor $\frac{dE^\sigma}{dB}$ causes the tilt to be spin-dependent and in the Zeeman limit of weak field $B \rightarrow 0$, we get $\Delta_\uparrow/\Delta_\downarrow = 7/9=78 \%$. With increasing field, the spin-dependence reduces (in the Paschen-Back limit $B \rightarrow \infty$ we have $\Delta_\uparrow/\Delta_\downarrow =1$) and for the magnetic field used in this work ($B \approx 210 \,\mathrm{G}$), we have $\Delta_{\uparrow}/\Delta_{\downarrow}=90.6\%$. The second factor $\partial_x B$ is spin-independent  and describes the magnetic field gradient, which we create with a single coil close to the atom cloud. The magnetic field along the 1D tubes is given by  $B(x) = B_0 + a (x-x_0) +b (x-x_0)^2$ plus higher orders which are negligible for our parameters. Here, $x_0$ is the center of the coil, $x$ is the relative distance of the atomic cloud, $a$ is the field gradient and $b$ is the field curvature. The coil has a diameter of $\SI{25}{\milli \meter}$, 20 windings and a mean distance to the atoms of $\SI{26.5}{\milli \meter}$. Currents up to $\SI{55}{\ampere}$ are applied. We note that the magnetic field generated by this configuration mainly possesses a large homogeneous contribution and a gradient part producing the linear potential. The weak field curvature part adds to the harmonic confinement of the lattice and dipole beams.

\subsection{Interaction averaging}

\label{sec:int_averaging}

\begin{figure}[!]
	\includegraphics[width=3.3in]{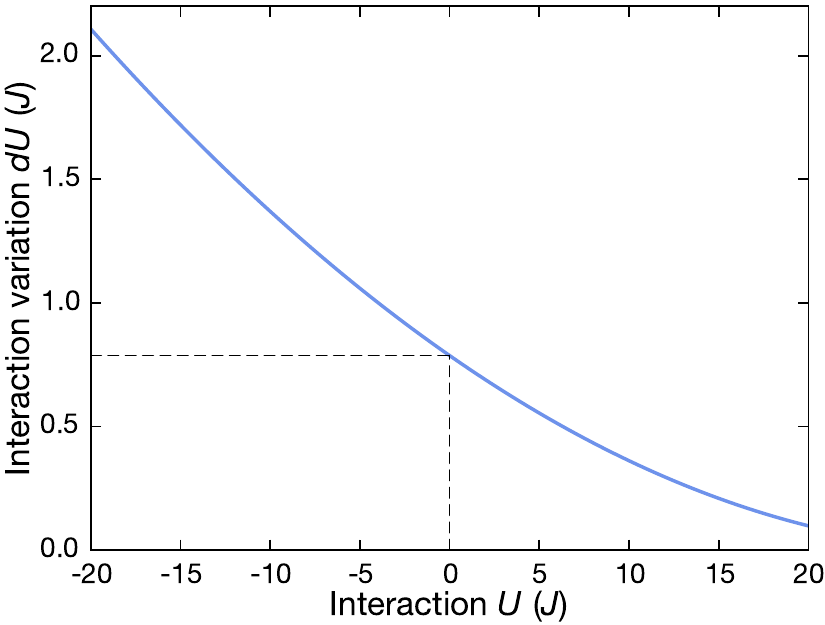}
	\caption{\textbf{Impact of interaction averaging.} Variation of the interaction strength across one tube with 290 lattice sites, a tilt of $\Delta_\downarrow/h=\SI{1.8}{\kilo\hertz}$ and tunneling rate $J/h=\SI{540}{\hertz}$. For the orthogonal lattices we use $55 E_r$.}
	\label{fig:Intaveraging}
\end{figure}

The magnetic field gradient used for generating the tilt causes a local variation of the total magnetic field. Since we use the total magnetic field to set the interaction strength with a Feshbach resonance,  the variation of the total magnetic field also induces a variation of the interaction strength over the length of a tube. From the typical center tube length of 290 lattice sites ($4\sigma$ width of the cloud) and the width of the Feshbach resonance we can calculate the impact of this averaging effect for a certain tilt and lattice configuration. Due to the Gaussian density distribution of the cloud, assuming $290$ sites as tube length overestimates the averaging effect and gives a crude upper bound. Fig.~\ref{fig:Intaveraging} shows the strength of the averaging effect $U_{\mathrm{var}}=U\pm dU$ as a function of the central interaction strength for a 1D system with $8E_r$ primary lattice depth. For shallower lattices this effect diminishes.  Note that while the Stark model exhibits a dynamical $U$ vs. $-U$ symmetry the interaction averaging slightly breaks this symmetry. It also underlines why the non-interacting data we show in this work is mostly taken with a spin-polarized sample. Even if the scattering length is set to zero via the Feshbach resonance,  small residual interactions remain due to the averaging.

\subsection{Calibration of parameters}

\subsubsection{Curvature $\alpha$ and trap frequency $\omega_h$}

Any non-linear correction to the linear on-site potential leads to a spectrum of Bloch oscillation frequencies in the system that are averaged over in the measurement. The non-linearity in our system is caused by the residual harmonic confinement, modelled as a quadratic correction term to the linear potential, $\Delta_\downarrow i + \alpha (i-i_0)^2$, where $i_0$ is the center of the lattice. The observed Bloch oscillation is then a sum of Bloch oscillations with frequencies ranging between $\Delta_\downarrow -2\alpha L/2$  and $\Delta_\downarrow + 2\alpha L/2$ with a step of $2\alpha$ and a system size of $L$ sites. In order to understand the result of such a sum, consider, for instance, a sum of sinusoidal oscillations,

\begin{equation}
\begin{split}
f(t)&=\sum_{i=-L/2}^{L/2}\cos(2\pi(\Delta_\downarrow+\alpha i)t) \\
&= \cos(2\pi\Delta_\downarrow t) \frac{\sin(2\pi (L+1)\alpha t)}{\sin(2\pi \alpha t)}.
\end{split}
\end{equation}

\begin{figure}[t!]
	\includegraphics[width=3.3in]{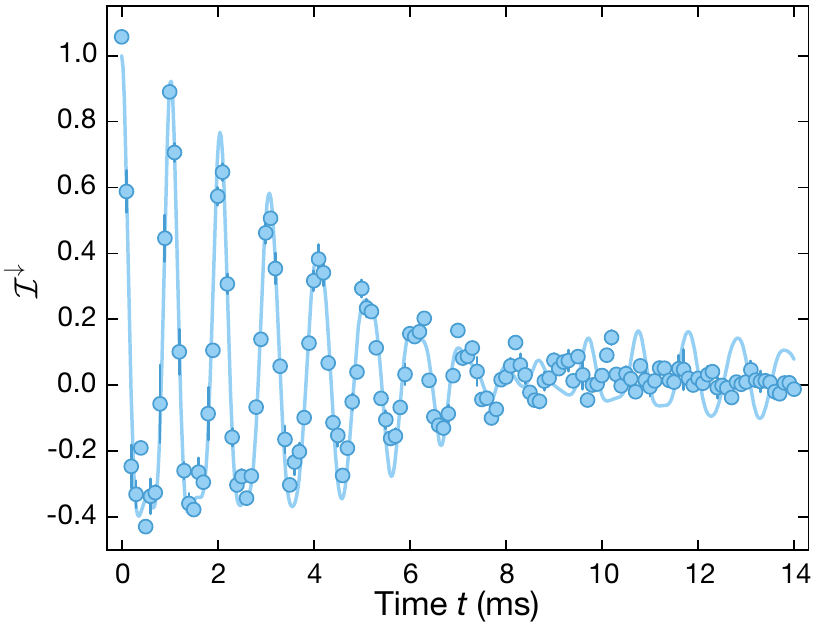}
	\caption{\textbf{Calibration of the harmonic confinement.} Imbalance $\mathcal{I}^\downarrow$ for a spin-polarized gas at $\Delta_\downarrow=1.8J$ and $J=h \cdot \SI{540}{\hertz}$. Each data point is averaged twice and error bars denote the SEM. The solid line is a fit to the data using an ED calculation, which includes the harmonic confinement. The resulting collapse time is $T_c=\SI{8}{\milli \second}$.}
	\label{fig:Confinementtrace}
\end{figure}

\noindent This is an oscillation at frequency $\Delta_\downarrow$ together with a beat note envelope at a frequency $(L+1)\alpha \approx L\alpha$ and nodes at $1/(2 L \alpha)$. The Bessel-type Bloch oscillations, which can be expressed as sum of few sinusoidal oscillations, would behave in a qualitatively similar manner. Therefore, we expect a collapse at time $T_c \approx 1/(2L\alpha)$, before the imbalance revives. We use numerical calculations of the imbalance time trace for a non-interacting system in a lattice of size $L=290 (20) d$ to determine the value of $\alpha$, as a fit parameter. Corresponding to an experimentally measured imbalance time trace $\mathcal{I}^\downarrow(t_j): j = 1, 2, \cdots, n$, where $n$ is the number of data points in time, we compute, numerically, the trace $\mathcal{I}^\downarrow_{\mathrm{num}}(t_j; J, \Delta_\downarrow, \alpha)$ and then minimize $\sum_j |\mathcal{I}^\downarrow(t_j)-\mathcal{I}^\downarrow_{\mathrm{num}}(t_j;  J, \Delta_\downarrow, \alpha)|^2$ over alpha to determine the fit value. The harmonic confinement is extracted in Fig.~\ref{fig:Confinementtrace}. We find a collapse time of $T_c = \SI{8}{\milli\second}$, corresponding to $\alpha=h \cdot \SI{0.216}{\hertz}$ and $\omega_h/ 2\pi = \sqrt{\frac{ \alpha \hbar}{m d^2 \pi}} = \SI{39}{\hertz}$. Due to the local nature of the dynamics in the Stark Hamiltonian, $\alpha$ is the important energy scale for the dynamics, characterizing the amount of curvature,  experienced by every single atom. In our system, the tilt is on the order of $\Delta_{\downarrow} \approx h\cdot \SI{1000}{\hertz}$ and the curvature is very weak ($\alpha/\Delta_{\downarrow} \approx 10^{-4}$). Theoretically, the imbalance oscillations should revive partially, but due to anharmonic confinement, residual onsite disorder and other dephasing mechanisms we cannot see such revivals. All these artifacts can affect the envelope of the Bloch oscillations in addition to the dephasing of the harmonic confinement and are also included in the extracted collapse time. Hence, extracting the harmonic confinement from the collapse time yields an upper bound for the true harmonic confinement.

\subsubsection{Spin-dependent tilt $\Delta_\sigma$}

\begin{figure}
	\includegraphics[width=3.3in]{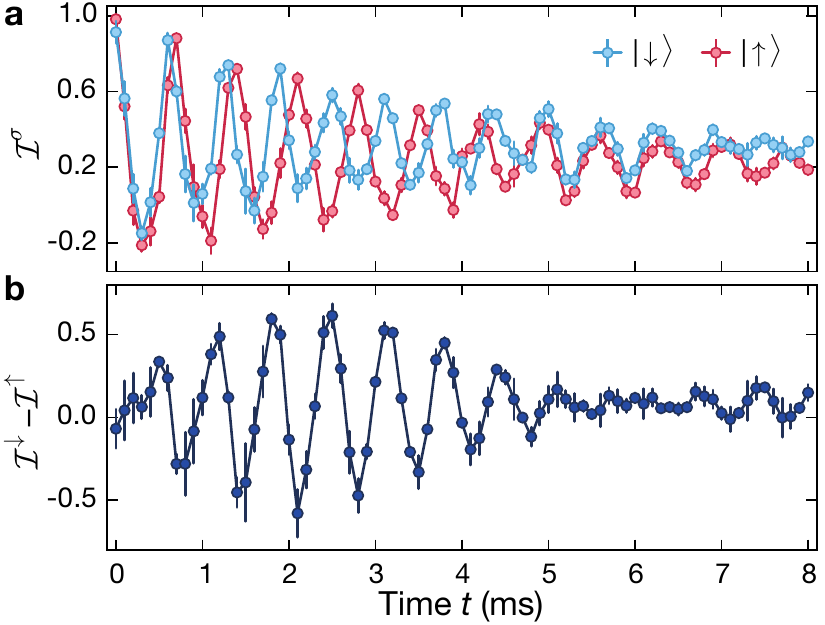}
	\caption{\textbf{Spin-resolved Bloch oscillations.} 
		\textbf{a} Typical calibration measurement of the tilt $\Delta_\sigma$ for both spin-components using the spin-resolved imbalance $\mathcal{I}^\sigma$. Here, $\Delta_\downarrow/h=\SI{1.60(1)}{\kilo\hertz}$ and we extract a frequency difference of $(\Delta_\downarrow-\Delta_\uparrow)/h=\SI{170(2)}{\hertz}$, which is in reasonable agreement with the calculated difference. Each data point is averaged four times and error bars denote the SEM. 
		\textbf{b} Imbalance difference between $\ket{\downarrow}$ and $\ket{\uparrow}$. The resulting pattern exhibits a beat note similar to the trigonometric identity $\cos(\omega_1t) - \cos(\omega_2t) = -2\sin((\omega_1+\omega_2)t/2) \sin((\omega_1-\omega_2)t/2)$.   }
	\label{fig_spinres_BO}
\end{figure}

The setup consists of one pair of coils in Helmholtz configuration to generate a homogeneous magnetic field $B_z$ along the vertical $z$ direction for controlling the interactions between the two spin states by a Feshbach resonance.  Additionally, a gradient coil is used to create a magnetic field $B_x$  consisting of a homogeneous field $B_{x0}$ and the field gradient $\frac{\mathrm{d}B_x}{\mathrm{d}x}$ along the $x$ direction. Therefore, the total field is $B_0 = \sqrt{B_x^2+B_z^2}$. Expanding the above expression up to first order, we get for the total field

\begin{equation}
\begin{split}
B_0(x) &= \sqrt{B_z^2 + B_x^2} = \sqrt{B_z^2 + \left(B_{x0} + x\frac{\mathrm{d}B_x}{\mathrm{d}x} \right)^2} \\
&\simeq B_z + \frac{B_{x0}^2}{2B_z} + \frac{B_{x0}}{B_z} \cdot x\frac{\mathrm{d}B_x}{\mathrm{d}x}.
\end{split}
\label{eq:magneticField}
\end{equation}

\noindent In the last step we used that $B_z$ is the strongest contribution such that the square root can be expanded up to first order and we neglected the term of the squared gradient. We note that the strength of the gradient is reduced by the vertical field component and amplified by the homogeneous horizontal field. It follows that the calibration of tilt and interactions has to be an iterative process, since these quantities strongly depend on each other. We determine the required vertical magnetic field $B_z$ in the presence of a current $I_G$ in the gradient coil in order to generate a fixed total homogeneous magnetic field $B_0$. For this sake we employ an RF sweep from $\ket{\downarrow}$ to $\ket{\uparrow}$, whose frequency is set to the value corresponding to $B_0$. We find the relation 

\begin{equation}
B_z(I_G) = B_0 - \frac{(aI_G)^2}{B_0} + bI_G \, ,
\end{equation}

\noindent with fit parameters $a$ and $b$. From Eq.~\eqref{eq:magneticField} it follows that $\Delta_\sigma \propto I_G^2$ where the proportionality constant depends on $B_z$. The current required to generate a certain tilt $\Delta_\sigma$ can thus be expressed as

\begin{equation}
I_G = c\sqrt{\Delta_\sigma \cdot B_z}
\label{eq:currentField}
\end{equation}

\noindent with constant $c$. We calibrate this fit parameter using single-particle Bloch oscillations and extract the oscillation frequency, set by the tilt $\Delta_\sigma$, with the analytical model using the first four oscillations to minimize effects of the damping. A typical calibration measurement is illustrated in Fig.~\ref{fig_spinres_BO} for a spin-mixture at $\Delta_\downarrow = h\cdot \SI{1.60(1)}{\kilo\hertz}$. We clearly see the different tilts in the oscillation frequency of the respective spin component. Finally, from Eq.~\eqref{eq:magneticField} and Eq.~\eqref{eq:currentField} we see that it requires an iterative adaption of the current and the vertical field, since they are strongly correlated. In the experiment we do two full iteration steps until the values sufficiently converge.

\begin{figure}[t!]
	\includegraphics[width=3.3in]{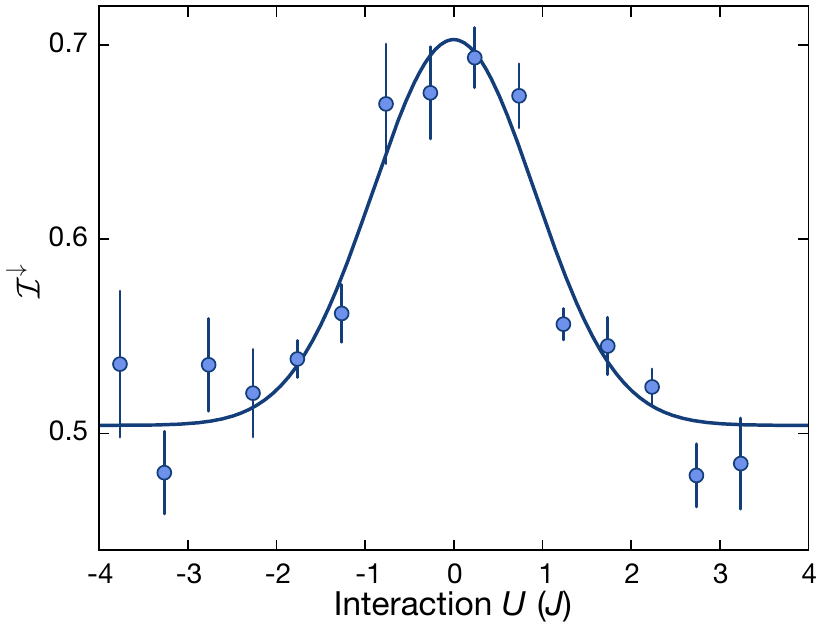}
	\caption{\textbf{Calibration of the zero-crossing.} Imbalance of one spin-component $\mathcal{I}^\downarrow$ versus interaction strength. We use a tilt $\Delta_\downarrow/h=\SI{1.2}{\kilo \hertz}$ and measure the imbalance after $t=h/\Delta_\downarrow$. The tunneling rate is $J/h=\SI{540}{\hertz}$. The solid line is a Gaussian fit to capture the peak of the imbalance, corresponding to the zero-crossing of the Feshbach resonance. Each data point consists of four independent measurements and error bars denote the SEM.}
	\label{fig:Uscan_Bloch}
\end{figure}

\subsubsection{Lattice depth}

All optical lattices are calibrated using Kapitza-Dirac scattering with a Bose-Einstein condensate of $^{87}\mathrm{Rb}$ and  the lattice depth calibration is then converted to $^{40}\mathrm{K}$. While this technique in principle also calibrates the tunneling $J$ in Eq.~\eqref{eq:Hamiltonian}, we determine the tunneling $J$ for the data in Fig.~\ref{fig2} and  Fig.~\ref{fig3} directly by using a fit of Eq.~\eqref{eq:Imbalance} to the short-time dynamics ($U=0J$, spin-polarized). We only use times $t \leq \SI{1.5} {\milli \second}$ such that the damping of the oscillations is negligible (the collapse time is $T_c=\SI{8}{\milli \second}$). For a set lattice depth of $8E_{rs}$ ($6E_{rs}$) the fit yields $J=h \cdot \SI{0.54(1)} {\kilo \hertz}$ ($J=h \cdot \SI{0.88(2)} {\kilo \hertz}$) and agrees in both cases well with the calculated tunnelling rate $J_{8E_{rs}}=h \cdot \SI{0.543}{\kilo \hertz}$ and  $J_{6E_{rs}}=h \cdot \SI{0.896}{\kilo \hertz}$. Note that the solid line in Fig.~\ref{fig3}b in the main text is a plot of Eq.~\eqref{eq:Imbalance_state}, where we use $J=h \cdot \SI{0.54(1)} {\kilo \hertz}$, obtained from the short time dynamics, without any additional free parameter. The excellent agreement of analytic prediction and data at late times emphasizes the accuracy of calibrating $J$ with the short time dynamics.

\subsubsection{Onsite interaction $U$}

The non-interacting point of the Feshbach resonance between the states $\ket{\uparrow}$ and $\ket{\downarrow}$ is calibrated with Bloch oscillations by taking advantage of the interaction-induced damping and the dynamical symmetry between repulsive and attractive interactions. For every tilt $\Delta_\downarrow$ we fix the time $t$ at $t=T_\downarrow=h/\Delta_\downarrow$, while scanning the Feshbach field. In Fig.~\ref{fig:Uscan_Bloch} we show a typical calibration measurement, where the zero-crossing of the Feshbach resonance is well detectable as the interaction strength, which has the largest imbalance. A finite interaction $U$ causes a strong damping, which decreases the imbalance. Since the magnetic field $B_0= 202.1 \, \mathrm{G}$ of the center of the Feshbach resonance is well known~\cite{regal_observation_2004}, the zero crossing is set by the width $w_{202}$ plus the magnetic field of the center $B_0$. We use the calibration of the zero crossing to determine a precise value for the width of the Feshbach resonance: $w_{202}=7.1(1) \, \mathrm{G}$, in agreement with the literature~\cite{schneider_fermionic_2012}. The same measurement was performed for the Feshbach resonance between $\ket{\downarrow}$ and $\ket{\rightarrow}$ centered at $B=224.2 \,\mathrm{G}$~\cite{regal_measurement_2003}, where we extract a width $w_{224}=7.4(1)\,\mathrm{G}$ in agreement with the literature~\cite{jordens_metallic_2010}.  
The characterization of the Feshbach resonance together with the calibration of the lattice depth yields a calibration for the onsite interaction $U$.

\subsection{Extraction of the interaction energy $U_{\mathrm{res}}$}

\begin{figure}[t!]
	\includegraphics[width=3.3in]{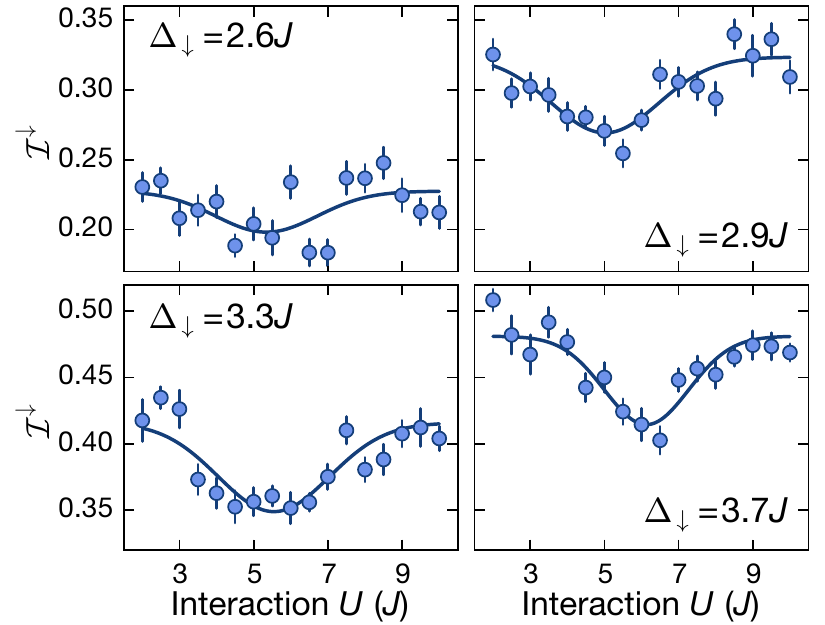}
	\caption{\textbf{Probing the interaction energy $U_{\mathrm{res}}$.} Interaction scan of the imbalance $\mathcal{I}^\downarrow$ of one spin component for different tilts $\Delta_\downarrow$. The solid line is a Gaussian fit $\mathcal{I}^\downarrow(U)=Ae^{-(U-U_{\mathrm{res}})^2/(2\sigma^2)}+C$ to the data extracting the interaction energy $U_{\mathrm{res}}$ for which the imbalance is minimal. The fit yields $U_{\mathrm{res}}=5.3(6)J$ $(\Delta_\downarrow=2.6J)$, $U_{\mathrm{res}}=5.0(2)J$ $(\Delta_\downarrow=2.9J)$, $U_{\mathrm{res}}=5.5(2)J$ $(\Delta_\downarrow=3.3J)$ and $U_{\mathrm{res}}=6.1(2)J$ $(\Delta_\downarrow=3.7J)$. Each data point is averaged three times over ten equally spaced times in a window between $170 \tau$ and $200 \tau$. Error bars denote the SEM. }
	\label{fig:IvsU_dip}
\end{figure}

In Fig.~\ref{fig3}e in the main text we discuss the resonant process connecting two singlons on even sites with a doublon via a second-order hopping process, with $U\simeq2\Delta_\downarrow$. In Fig.~\ref{fig:IvsU_dip}, we show how the interaction energy $U_{\mathrm{res}}$ of the resonance for different tilts $\Delta_\downarrow$ was extracted using a Gaussian fit $f(x)=Ae^{-(x-x_c)^2/(2\sigma^2)}+C$ to locate the minimum of the imbalance $\mathcal{I}^\downarrow$. Note that the naive expectation for the interaction energy $U_{\mathrm{res}}=2\Delta_\downarrow$ does not apply here, because the resonance is of second order and therefore the interaction energy is renormalized such that we expect $U_{\mathrm{res}}+8J^2/(3\Delta_\downarrow)=2 \Delta_\downarrow$ up to second-order perturbation theory.

\subsection{Details on numerical methods}

We use an exact diagonalization technique to simulate the long-time dynamics. The dimension of the subspace of $N_\sigma$ spin $\sigma$ atoms on $L$ lattice sites is $d_{\sigma} = \binom{L}{N_{\sigma}}$. The state $\psi$ is a $d_{\uparrow}d_{\downarrow}$ dimensional vector. The total dimension of the Hilbert space is $d_{\uparrow}d_{\downarrow}$ --- $H$ is a $d_{\uparrow}d_{\downarrow}\times d_{\uparrow}d_{\downarrow}$ matrix. For $L=12, N_{\sigma}=3$, this dimension is $220^2 = 48400$; for $L=16, N_{\sigma}=4$, it is $1820^2 = 3312400$ and for $L=20, N_{\sigma}=5$, it is $15504^2 = 240374016$. At $L=12$, the Hamiltonian already consists of $48400^2$ floating point numbers, occupying up to $75 \,\mathrm{GB}$ of RAM. We therefore use the following method for the computation, which enables us to go up to $L=20$ before using sparse matrices or a Lanczos algorithm. 
Below, we describe the construction of the basis, the Hamiltonian and time evolution.

\subsubsection{Basis construction} Accounting for atom number conservation in both the spins, as mentioned before, we are working in a subspace of dimension $d_{\up}\times d_{\dn}$. This is a system of $N_{\up}+N_{\dn}$ fermions with a total of $2L$ fermionic modes represented by the creation operators $\hat c_{1, \up}^{\dagger},\hat c_{2, \up}^{\dagger}, \cdots,\hat c_{L, \up}^{\dagger}, \hat c_{1, \dn}^{\dagger},\hat c_{2, \dn}^{\dagger}, \cdots, \hat c_{L, \dn}^{\dagger}$. A typical number state can be written as $\hat c_{i_1, \up}^{\dagger} \hat c_{i_2, \up}^{\dagger}\cdots \hat c_{i_{N_{\up}}, \up}^{\dagger} \hat c_{j_1, \dn}^{\dagger} \hat c_{j_2, \dn}^{\dagger}\cdots \hat c_{j_{N_{\dn}}, \dn}^{\dagger} \ket{0} $, where $\{i_1, \cdots, i_{N_{\up}}\}$ and  $\{j_1, \cdots, j_{N_{\dn}}\}$ are subsets (not necessarily disjoint) of $\{1, 2, \cdots, L\}$. We construct a canonical representation of this state by ordering the operators such that $i_1 < i_2 <  \cdots < i_{N_{\up}}$  and $j_1 < j_2 <  \cdots < j_{N_{\dn}}$. This state can be represented by the pair of tuples $((i_1, i_2, \cdots, i_{N_{\up}}), (j_1, j_2, \cdots, j_{N_{\dn}}))$. Next we order the tuples $\{(i_1, i_2, \cdots, i_{N_{\up}} )\}$ lexicographically to construct a list of tuples $\mathcal V_{\up}$ and $\mathcal V_{\dn}$. The full basis would then be $\mathcal V_{\up} \times \mathcal V_{\dn}$. In this basis, the non-interacting part of the Hamiltonian remains separable and we make use of this property to optimize the time and memory consumption.

\paragraph{Off-diagonal elements of the Hamiltonian:} Note that a typical hopping term in the Hamiltonian corresponding to spin $\up$ atoms not only leaves the spin $\dn$ part of a basis element unchanged, but also maintains the sign of the state with the trivial exception of the boundary hopping (e.g., $\hat c_{1, \up}^{\dagger} \hat c_{L, \up}$).
The hopping matrix can, therefore, be written as $\hat H^{\mathrm{hop}}_{\up}\otimes 1 + 1 \otimes \hat H^{\mathrm{hop}}_{\dn}$ where $\hat H^{\mathrm{hop}}_{\sigma}=\sum_i \hat c_{i,\sigma}^\dagger \hat c_{i+1,\sigma}+\mathrm{h.c.}$ is the $d_{\sigma}\times d_{\sigma}$ matrix corresponding to the hopping of spin $\sigma$ atoms, acting on $\text{span}(\mathcal V_{\sigma})$. We construct $\hat H^{\mathrm{hop}}_{\up}$ and $ \hat H^{\mathrm{hop}}_{\dn}$ separately.
These two matrices are \textit{small}, in the sense that their dimensions are $d_{\up}$ and $d_{\dn}$, much smaller than the full Hilbert space $d_{\up}\times d_{\dn}$. These matrices can be stored in as dense matrices even when $L=20$ and $N_{\sigma}=5$.

\paragraph{Diagonal elements:} We represent the potential of a spin $\sigma$ atom at site $i$ by $V_{i, \sigma}$. We store \textit{only} the diagonal entries of the Hamiltonian in a matrix $V$ of size $d_{\up} \times d_{\dn}$. The matrix element $V_{\alpha\beta}$ is the energy of the basis element corresponding to $\alpha$-th tuple in $\mathcal V_{\up}$ and $\beta$-th tuple in $\mathcal V_{\dn}$. If this basis element is $((i_1, i_2, \cdots, i_{N_{\up}}), (j_1, j_2, \cdots, j_{N_{\dn}}))$, the energy is

\begin{equation*}
V_{\alpha\beta} = \sum_{k=1}^{N_{\up}}V_{ i_k, \up}+\sum_{k=1}^{N_{\dn}}V_{ j_k, \dn} + U|\{i_1, \cdots, i_{N_{\up}}\}\cap \{j_1, \cdots, j_{N_{\dn}}\}|
\end{equation*}  
Here, $|\{i_1, \cdots, i_{N_{\up}}\}\cap \{j_1, \cdots, j_{N_{\dn}}\}|$ is the number of elements in the intersection of $\{i_1, \cdots, i_{N_{\up}}\}$ and $\{j_1, \cdots, j_{N_{\dn}}\}$. This is the number of doublons in the state and the last term in the above equation corresponds to the Hubbard interaction.
\subsubsection{Intermediate time evolution}

We define a $d_{\uparrow}\times d_{\downarrow}$ matrix $M^{(\psi)}$, storing the state $\psi$, whose $\alpha\beta$-th element is $M^{(\psi)}_{\alpha\beta}=\langle \alpha, \beta|\psi\rangle $. Here, $|\alpha\beta\rangle$ is the basis element with indices $(\alpha, \beta)$ in $\mathcal V_{\up}\times \mathcal V_{\dn}$. The rows of $M^{(\psi)}$ correspond to spin $\uparrow$ and columns correspond to spin $\downarrow$. 
With this setting, the state is $M^{(\psi)}$ and the Hamiltonian, represented by the triplet $\{\hat H^{\mathrm{hop}}_{\uparrow},\hat H^{\mathrm{hop}}_{\downarrow},\hat V\}$, all of which are $d_{\uparrow}\times d_{\downarrow}$, $d_{\downarrow}\times d_{\downarrow}$ or $d_{\uparrow}\times d_{\uparrow}$ matrices.  Therefore it is convenient to work in this picture rather than use the full Hamiltonian which is much bigger. The Schr\"odinger equation in this representation is given by

\begin{equation}\label{time_evolution}
\dot{M^{(\psi)}} = -i \hat H^{\mathrm{hop}}_{\uparrow}M^{(\psi)} - i M^{(\psi)} \hat H^{\mathrm{hop}}_{\downarrow} -i \hat V\circ M^{(\psi)}
\end{equation} 
Here, $\circ$ represents element-by-element multiplication, known as Hadamard product. To see that this is the correct equation of time evolution, consider the Schr\"odinger equation in the standard representation
\begin{equation}\label{time_evolution_original}
\dot{\psi} = -i \hat H^{\mathrm{hop}}_{\uparrow}\otimes 1 \psi - i 1\otimes \hat H^{\mathrm{hop}}_{\downarrow} \psi -i \hat H^{diag} \psi
\end{equation} 
Here, $\hat H^{\mathrm{diag}}$ is a $d_{\up}d_{\dn} \times d_{\up}d_{\dn}$ diagonal matrix consisting of the elements in $\hat V$.
The first term in the above equation reads $\hat H^{\mathrm{hop}}_{\uparrow}\otimes 1 \psi = \sum_{\alpha, \beta}\sum_{\gamma} \hat H^{\mathrm{hop}}_{\uparrow \alpha \gamma} M^{(\psi)}_{\gamma\beta} \ket{\alpha\beta} = \sum_{\alpha, \beta} (\hat H^{\mathrm{hop}}_{\uparrow}M^{(\psi)})_{\alpha\beta}\ket{\alpha\beta}$. The second terms reads  $1\otimes \hat H^{\mathrm{hop}}_{\downarrow} \psi = \sum_{\alpha, \beta}\sum_{\gamma} \hat H^{\mathrm{hop}}_{\downarrow \beta \gamma} M^{(\psi)}_{\alpha\gamma}\ket{\alpha\beta} = \sum_{\alpha, \beta} (M^{(\psi)} \hat H^{\mathrm{hop}}_{\downarrow})_{\alpha\beta}\ket{\alpha\beta}$. Thus the first term corresponds to a \textit{left} multiplication of $M^{(\psi)}$ by $\hat H^{\mathrm{hop}}_{\uparrow}$ and the second term corresponds to a \textit{right} multiplication by $\hat H^{\mathrm{hop}}_{\downarrow}$. The third term reads 
$\hat H^{\mathrm{diag}} \psi = \sum_{\alpha, \beta} \hat V_{\alpha\beta} M^{(\psi)}_{\alpha\beta}\ket{\alpha\beta}$; this corresponds to a \textit{term-by-term} multiplication of $M^{(\psi)}$ by $\hat V$. 
Using Trotter-Suzuki decomposition to solve Eq.~\eqref{time_evolution} yields

\begin{equation}\label{TS_timestep}
M^{(\psi)}(t+\delta t) \approx e^{-i\delta t \circ \hat V}\circ e^{-i \delta t \hat H^{\mathrm{hop}}_{\uparrow}}M^{(\psi)}(t) e^{-i \delta t \hat H^{\mathrm{hop}}_{\downarrow}}
\end{equation} 
Here, $e^{-i\delta t \circ \hat  V}$ is an element-by-element exponentiation.

\begin{figure*}[t!]
	\includegraphics[width=\textwidth]{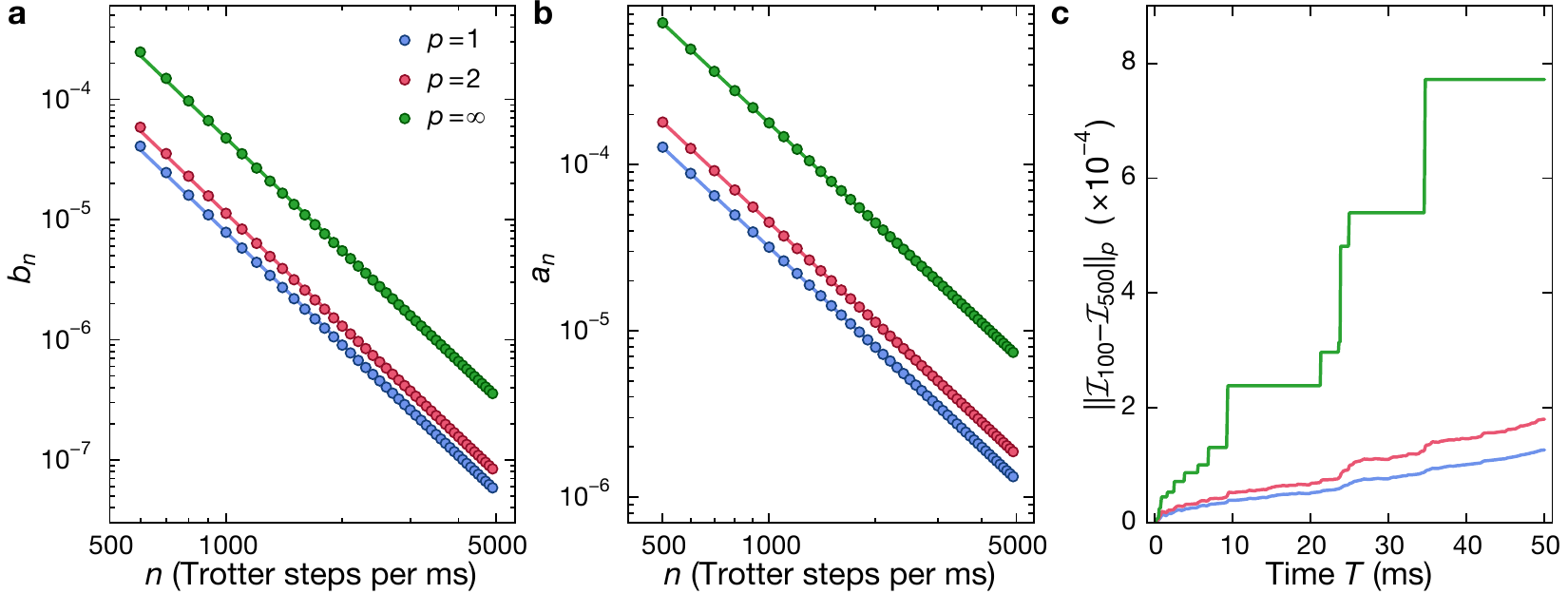}
	\caption{\textbf{Error estimates in Eq.~\eqref{TS_timestep}.} 
		In panels \textbf{a} and \textbf{b}, the $p-$norms are calculated for $T=\SI{100}{\milli \second}$ \textbf{a} Computation of $b_n = ||\mathcal I_n -\mathcal I_{n+r}||_p$ with $r=100$ for $L=8, N_{\sigma}=2$. We show $b_n$ for $p=1, 2$ and $\infty$. The solid lines are power law fits, and the corresponding exponent is about $-3.09$. 
		\textbf{b} Computation of $a_n = ||\mathcal{I} -\mathcal I_{n}||_p$ for a system with $L=8, N_{\sigma}=2$. The solid lines correspond to the respective estimates of $\mathcal I(t)$ given by Eq.~\eqref{err_estimate}. The three colors represent $p=1, 2$ and $\infty$. 
		\textbf{c} The growth of the errors $||\mathcal I_{100}-\mathcal I_{500}||_p$ in time for $p=1, 2$ and $\infty$, for $L=8$ and $N_{\sigma}=2$.}
	\label{fig:numerics_1}
\end{figure*}

\subsubsection{Error estimates}

The key problem here is to estimate the error accumulated due to the Trotter-Suzuki decomposition. We consider $n$ trotter steps per Bloch period, that is, $\delta t =\frac{1}{n\Delta}$. We consider time evolutions up to time $T$. If $\mathcal I_n(t)$ is the imbalance computed using $n$ trotter steps per Bloch period, our objective is to estimate $||\mathcal I - \mathcal I_n||$ where $\mathcal I(t)$ is the imbalance at $n=\infty$, which we can calculate for small system sizes using Eq.~\eqref{time_evolution_original}, by exponentiating the full Hamiltonian. We use the standard $\mathcal L^p$-norm, i.e., $||\mathcal I_n - \mathcal I||_p = \left(\int_0^T |\mathcal I_n(t) - \mathcal I(t)|^pdt\right)^{\frac{1}{p}}$ for $p=1, 2$ and $\infty$. In the latter case, $||\mathcal I_n -\mathcal I||_{\infty} = \mathrm{max}(|\mathcal I(t)-\mathcal I_n(t)|)$. Intuitively, $p=1$ corresponds to the "average case" distance between $\mathcal I$ and $\mathcal I_{\infty}$ and $p=\infty$ represents the "worst case" distance.

We numerically show that $a_n = ||\mathcal I - \mathcal I_n|| = O\left(\frac{1}{n^2}\right)$. In other words, $||\mathcal I - \mathcal I_n|| \rightarrow 0$ as $1/n^2$. To see this, let us consider the sequence $b_n = ||\mathcal I_{n+r} - \mathcal I_n||$, for a fixed $k$. Fig.~\ref{fig:numerics_1}a shows that $b_n = O\left(\frac{1}{n^3}\right)$. Moreover, from triangle inequality, $|a_{n+r}-a_n|\leq b_n$ and therefore, $a_n = O\left(\mathrm{cusum}\left(\frac{1}{n^3}\right)\right) = O\left(\frac{1}{n^2}\right)$. Here, $\mathrm{cusum}(x)$ is the cumulative sum. Thus, for large $n$ and some $k$, we can assume that $a_{n} \sim k^2 a_{kn}$  and it follows from the triangle inequality, $a_n -a_{kn} \leq ||\mathcal I_{kn} - \mathcal I_n||$ that $a_n \approx  \frac{k^2}{k^2-1}||\mathcal I_{kn} - \mathcal I_n||$. The RHS of  last inequality can be computed numerically. Thus, we obtain

\begin{equation}\label{err_estimate}
||\mathcal I_m -\mathcal I|| \approx \frac{k^2 n^2}{(k^2-1)m^2}||\mathcal I_{kn} -\mathcal I_n||
\end{equation}

We use $n=100$ and $k=5$ in Fig.~\ref{fig:numerics_1}b. We use the above expression to estimate $a_n = ||\mathcal I - \mathcal I_n||$, for $L=12$ and higher, and choose $n$ such that $a_n \leq 10^{-3}$.

\begin{figure*}
	\includegraphics[width=\textwidth]{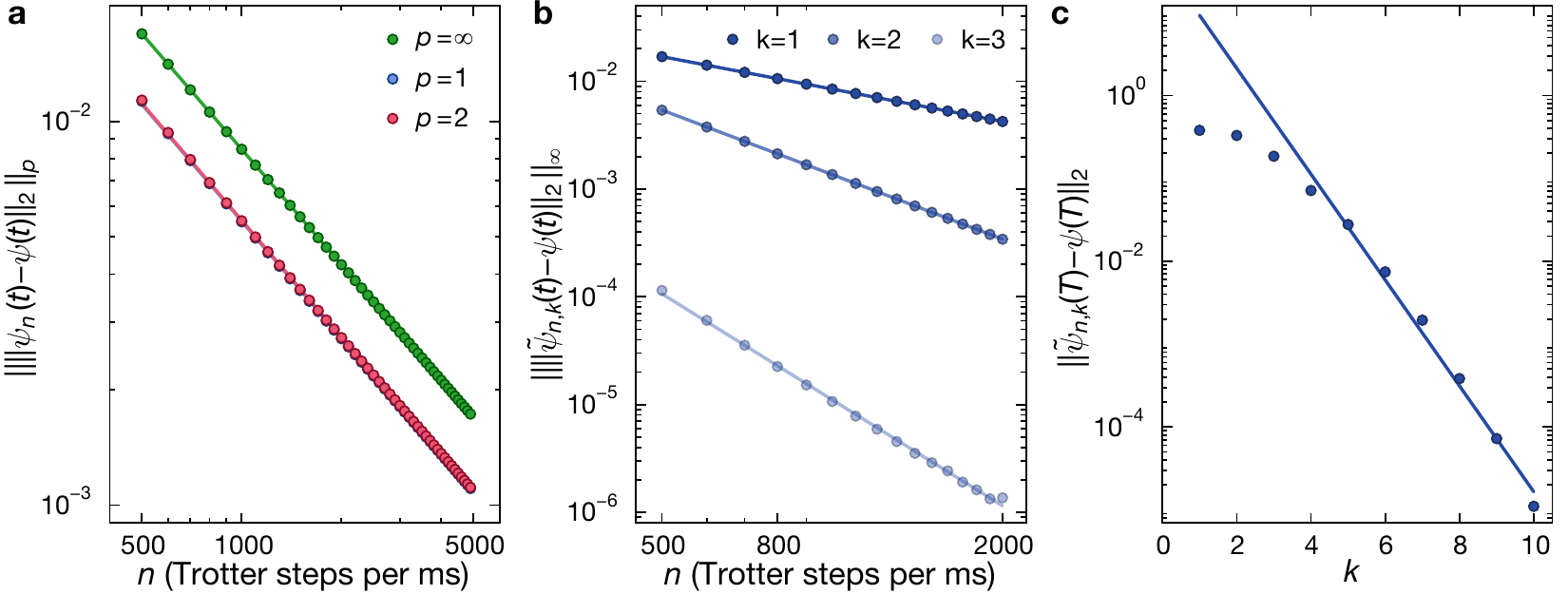}
	\caption{\textbf{General error analysis.} 
		\textbf{a} $||||\psi_n(t)-\psi(t)||_2||_p$ for a system with $L=8, N_{\sigma}=2$, indicating a convergence rate of $\frac{1}{n}$. 
		\textbf{b} $||||\tilde{\psi}_{n,k}(t)-\psi(t)||_2||_{\infty}$ for different $k=1,2,3$, system size $L=8$ and $ N_{\sigma}=2$ with $T=\SI{100}{\milli \second}$. $k=2$ corresponds to the elimination of the first order error following Eq.~\eqref{k_2_reduction}. The $k=3$ curve corresponds to an elimination of the first two orders in error. The straight lines are power law fits with exponents $-0.99$, $-1.99$ and $-3.26$ respectively for  $k=1, 2$ and $3$, indicating a convergence rate of $\frac{1}{n^k}$. 
		\textbf{c} $||\tilde{\psi}_{n,k}(T)-\psi(T)||_2$ at $T=\SI{100}{\milli \second}$ for a system with $L=8, N_{\sigma}=2$ and $n_1=40$ for various $k$, indicating a convergence rate of $\frac{1}{n_1^k}$.}
	\label{fig:numerics_2}
\end{figure*}

\textit{General error analysis} In the previous section, the error accumulated due to the Trotter-Suzuki approximation was analysed using the deviations in the imbalance as the figure of merit. While this approach is relevant for our purpose, the deviations in the state vector itself would be relevant in a more general context. Indeed, if $\psi(t)$ is the many body state vector at time $t$ and $\psi_n(t)$ is the state vector computed using a Trotter-Suzuki approximation using $n$ Trotter steps per Bloch period, the deviation $||||\psi_n(t)-\psi(t)||_2||_p$ can be related to the deviation in any arbitrary observable $\hat{O}$ (Note that we use the standard 2-norm to quantify the distance between $\psi(t)$ and $\psi_n(t)$ at a given time $t$ and then use a $p-$norm to quantify the overall deviation). For instance, consider $\langle \psi|\hat{O}|\psi\rangle-\langle \psi_n|\hat{O}|\psi_n\rangle=\langle \psi|\hat{O}|\psi\rangle-\langle \psi|\hat{O}|\psi_n\rangle+\langle \psi|\hat{O}|\psi_n\rangle-\langle \psi_n|\hat{O}|\psi_n\rangle= \langle \psi-\psi_n|\hat{O}|\psi\rangle+\langle \psi_n|\hat{O}|\psi-\psi_n\rangle \leq 2||\hat{O}||||\psi-\psi_n||_2$. Here, $||\hat{O}||$ is the operator norm, i.e., the largest singular value of $\hat{O}$, assuming $\hat{O}$ is finite-dimensional. Thus, the deviation $||\langle\psi_n(t)|\hat{O}|\psi_n(t)\rangle-\langle\psi(t)|\hat{O}|\psi(t)\rangle||_p$ can be estimated, loosely, using $||||\psi_n(t)-\psi(t)||_2||_p$. In Fig.~\ref{fig:numerics_2}a we show that $||||\psi_n(t)-\psi(t)||_2||_p\sim \frac{1}{n}$ for a system with $L=8$ sites and $N_{\sigma}=2$. It is interesting to note that the imbalance converges much faster $\sim \frac{1}{n^2}$.

\subsubsection{Long-time evolution}
At the outset it appears that by making a Trotter-Suzuki decomposition, we lose the logarithmic scaling of complexity in time of the scaling-and-squaring procedure of matrix exponentiation. That is, for a long time $T$, the unitary $e^{-i \hat H T}$ can be computed by scaling $T$ to $T/2^n$ for some integer $n$, computing $e^{-i \hat H T/2^n}\approx 1-i \hat H T/2^n - \hat H^2 /2 T^2/4^n$ and squaring it repeatedly, $n$ times. The complexity of this procedure is linear in $n$. For a fixed tolerance, it is logarithmic in $T$, enabling a computation of very long-time dynamics. Although it appears that we lose this advantage while using the Trotter-Suzuki decomposition, we show below that the scaling can be improved, asymptotically. 

In Fig.~\ref{fig:numerics_1}c  we show how the error of the computation increases with time for a fixed trotter step, suggesting at least a linear growth in $T$. Moreover, for fixed trotter step, the computational time also grows linearly in $T$. Thus, for a fixed tolerance, the computational time grows at least quadratically in $T$. We show below that this can be improved to a linear scaling in $T$.
The idea is to reduce the error in the computation using an elimination technique so that it scales down faster. Let $T\gg 1/\Delta$ be a long time up to which we intend to compute the evolution the system. That is, we want to compute $\psi (T)$. Let us suppose that we computed $\psi (T)$ twice, using the above described procedure, once using $\delta t = 1/(n\Delta)$ and the second time using  $\delta t = 1/((n+1)\Delta)$ and obtained two state vectors $\psi_n(T)$ and $\psi_{n+1}(T)$. From the above considerations, we know that $||\psi(T)-\psi_{n}(T)||=\kappa /n + O(1/n^2)$ for some $\kappa$. We consider a linear combination of $\psi_n(T)$ and $\psi_{n+1}(T)$:

\begin{equation}\label{k_2_reduction}
\tilde{\psi}_{n, 2}(T)= (n+1)\psi_{n+1}(T)-n\psi_n(T)
\end{equation}
$\tilde{\psi}_{n, 2}$ is an attempt to eliminate the first order term in the error and therefore, we expect the error scaling to be lower for this state. Indeed, as shown in Fig.~\ref{fig:numerics_2}b,  $||\tilde{\psi}_{n, 2}-\psi(T)||\sim 1/n^2$.
We may consider a general procedure to eliminate the higher order error terms. We pick $k$ integers $n_1, \cdots, n_k$ and compute $\psi_{n_1}(T), \cdots, \psi_{n_k}(T)$ independently and eliminate the first $k-1$ orders of error. This is done using the Vandermonde matrix

\begin{equation*}
W_k= \left(\begin{array}{cccc}
1 & 1/n_1 & \cdots & 1/n_1^{k-1}\\
1 & 1/n_2 & \cdots & 1/n_2^{k-1}\\
\vdots & \vdots & \ddots & \vdots \\
1 & 1/n_k & \cdots & 1/n_{k}^{k-1}\\
\end{array}\right)
\end{equation*}

We can eliminate the errors using the expression $\tilde{\psi}_{n_1, k}=\sum_j(W_{k}^{-1})_{1j}\psi_{n_j}(T)$. Fig.~\ref{fig:numerics_2}c shows that the error in $\tilde{\psi}_{n_1, k}$ scales down as $\sim 1/n_1^k$.


\begin{thebibliography}{97}
\bibitem [{imb()}]{imbalance}%
	\BibitemOpen
	\href@noop {} {}\bibinfo {note} {Note, imbalance is defined as a charge
		imbalance between even and odd lattice sites in our system and does not probe
		spin imbalances. The magnetization of systems is conserved during the
		evolution and it is equal to zero at all times.}\BibitemShut {Stop}%
\end{thebibliography}
\end{document}